\documentclass[conference,letterpaper,10pt]{IEEEtran}
\usepackage{graphicx,cite}
\usepackage{caption}
\usepackage{subcaption}
\usepackage{amssymb,amsmath,amsbsy}

\usepackage{array}

\usepackage{balance}

\usepackage[hyphens]{url}
\usepackage{hyperref}

\usepackage{multirow}

\begin{document}
\graphicspath{{figures/}}

\title{RepNet: Cutting Tail Latency in Data Center \\Networks with Flow Replication}


\author{\IEEEauthorblockN{Shuhao Liu\IEEEauthorrefmark{1}\IEEEauthorrefmark{3}, Wei Bai\IEEEauthorrefmark{2}, Hong Xu\IEEEauthorrefmark{1}, Kai Chen\IEEEauthorrefmark{2}, Zhiping Cai\IEEEauthorrefmark{3}}
\IEEEauthorblockA{ 
\IEEEauthorrefmark{1}Department of Computer Science, City University of Hong Kong
}
\IEEEauthorblockA{ 
\IEEEauthorrefmark{2}Department of Computer Science and Engineering, HKUST
}
\IEEEauthorblockA{ 
\IEEEauthorrefmark{3}College of Computer, National University of Defence Technology
}
}

\maketitle

\begin{abstract}

Data center networks need to provide low latency, especially at the tail, as demanded by many interactive applications. To improve tail latency, existing approaches require modifications to switch hardware and/or end-host operating systems, making them difficult to be deployed. We present the design, implementation, and evaluation of RepNet, an application layer transport that can be deployed today. RepNet exploits the fact that only a few paths among many are congested at any moment in the network, and applies simple flow replication to mice flows to opportunistically use the less congested path. RepNet has two designs for flow replication: (1) RepSYN, which only replicates SYN packets and uses the first connection that finishes TCP handshaking for data transmission, and (2) RepFlow which replicates the entire mice flow. We implement RepNet on {\tt node.js}, one of the most commonly used platforms for networked interactive applications. {\tt node}'s single threaded event-loop and non-blocking I/O make flow replication highly efficient. Performance evaluation on a real network testbed and in Mininet reveals that RepNet is able to reduce the tail latency of mice flows, as well as application completion times, by more than 50\%.

\end{abstract}

\section{Introduction}
\label{sec:intro}

Data center networks are increasingly tasked to provide low latency for many interactive applications they support \cite{AGMP10,ZDMB12,AYSK13}. Low tail latency (e.g. 99th or 99.9th percentile) is especially important for these applications, since a request's completion depends on all (or most) of the responses from many worker machines \cite{DB13}. Unfortunately current data center networks are not up to this task: Many report that the tail latency of short TCP flows can be more than 10x worse than the average in production networks, even when the network is lightly loaded \cite{AYSK13,XMNB13,ZDMB12}.

The main reason for long tail latency is two-fold. First, elephant and mice flows co-exist in data center networks. While most flows are mice with less than say 100~KB, most bytes are in fact from a few elephants \cite{GHJK09,AGMP10,KSGP09}. Thus mice flows are often queued behind bursts of packets from elephants in switches, resulting in long queueing delay and flow completion time (FCT). Second, and more importantly, despite the recent progress in high bisection bandwidth topologies \cite{ALV08,GWTS09,GWTS08,MPFH09}, the core part of the network is still over-subscribed in most production settings for cost and scalability reasons. This makes congestion more likely to happen in the network rather than at the edge.

The problem has attracted much attention recently in our community \cite{LXC13}. Loosely speaking, existing work reduces the tail latency by: (1) reducing the queue length, such as DCTCP \cite{AGMP10} and HULL \cite{AKEP12}; (2) prioritizing mice flows, such as D$^3$ \cite{WBKR11}, PDQ \cite{HCG12}, pFabric \cite{AYSK13}, and PIAS \cite{BCCH14}; and (3) engineering better multi-path schemes, such as DeTail \cite{ZDMB12}, DRB \cite{CXYG13}, and Expeditus \cite{WX14}. While effective, they require changes to switches and/or end-hosts, and face significant deployment challenges. Thus there is a growing need for an application layer solution that provides immediate latency gains without an infrastructure overhaul.

To this end, we introduce RepNet, a low latency transport at the application layer that can be readily deployed in current infrastructures. RepNet is based on the simple idea of flow replication to reap the path diversity gains. Due to the random nature of the traffic and hash based load balancing, while some paths may be heavily utilized, many other paths are uncongested in a large-scale network. Thus, we can replicate a mice flow by creating another TCP connection, and it is highly unlikely that both flows experience congestion and long queueing delay.
Additionally, flow replication is \emph{orthogonal} to all TCP-friendly proposals in the literature. Thus it can be used together with schemes such as DCTCP \cite{AGMP10} and pFabric \cite{AYSK13}, providing even more benefit in reducing latency.


In this paper we present the design, implementation, and evaluation of RepNet based on flow replication, whose benefit has only been theoretically established \cite{XL14}. We make three concrete contributions. 

First, we propose RepSYN which only replicates the SYN packets to reduce the overhead and performance penalty of flow replication. Directly realizing flow replication means we shall replicate each and every packet of a flow on a second TCP connection, which is proposed and studied in the RepFlow paper \cite{XL14}. Yet an astute reader might be concerned about the overhead of using RepFlow, especially in incast scenarios where many senders transmit at the same time to a common destination causing throughput collapse \cite{VPSK09}. RepFlow potentially aggravates the incast problem. To address this, RepSYN only replicates the SYN packet on the second TCP connection, and uses the connection that finishes handshaking first for data transmission.

Second, we implement RepNet---with both RepFlow and RepSYN---on {\tt node.js} \cite{node-website} as a transport module that can be directly used by existing applications running in data centers. {\tt node.js} (or simply {\tt node}) is a server-side JavaScript platform that uses a single-threaded event-loop with a non-blocking I/O model, which makes it ideal for replicating TCP flows without much performance overhead. Moreover, {\tt node} is widely used for developing the back-end of large-scale interactive applications in production systems at LinkedIn \cite{nodeatlinkedin}, Yahoo!, Microsoft, etc.\footnote{\url{http://nodejs.org/industry/}} RepNet on {\tt node} potentially provides immediate latency benefit for a large number of these applications with minimal code change. 

Our third contribution is a comprehensive performance evaluation of RepNet on a small scale leaf-spine testbed as well as a 6-pod fat-tree in Mininet \cite{HHJL12}, using an empirical flow size distribution from a production network. Our evaluation shows that, both RepFlow and RepSYN significantly reduces the tail latency of mice flows, especially under high loads, by more than 50\%. RepSYN is less effective compared with RepFlow in most cases, but it remains beneficial in incast scenarios where RepFlow suffers from performance degradation. We further implement a bucket sort application using RepNet, and observe that both RepFlow and RepSYN improves the application level completion times by around 50\%. The implementation code \cite{repnet}, and scripts used for performance evaluation, are available online \cite{repnet_exp}. We are in the process of making RepNet available as an NPM (Node Package Manager) module for the {\tt node} user community.




\section{Motivation and Design }
\label{sec:motivation_design}



Let us start by motivating the idea of flow replication to reduce latency in data center networks, followed by the high-level design of {RepNet} including both RepFlow \cite{XL14} and RepSYN.

\subsection{Motivation}
\label{sec:motivation}

Today's data center networks are constructed following Clos topologies such as fat-tree \cite{ALV08}. In these topologies, many paths of equal distance exist between a pair of hosts. Equal-cost multi-path routing, or ECMP, is used to perform flow-level load balancing \cite{ecmp-rfc} that routes packets based on the hash value of the five-tuple in the packet header. Due to the randomness of traffic and ECMP, congestion happens randomly in some paths of the network, while many others are not congested at all.

\begin{figure}[hbt]
	\vspace{-0.0cm} \centering
	\includegraphics[width=0.44\textwidth]{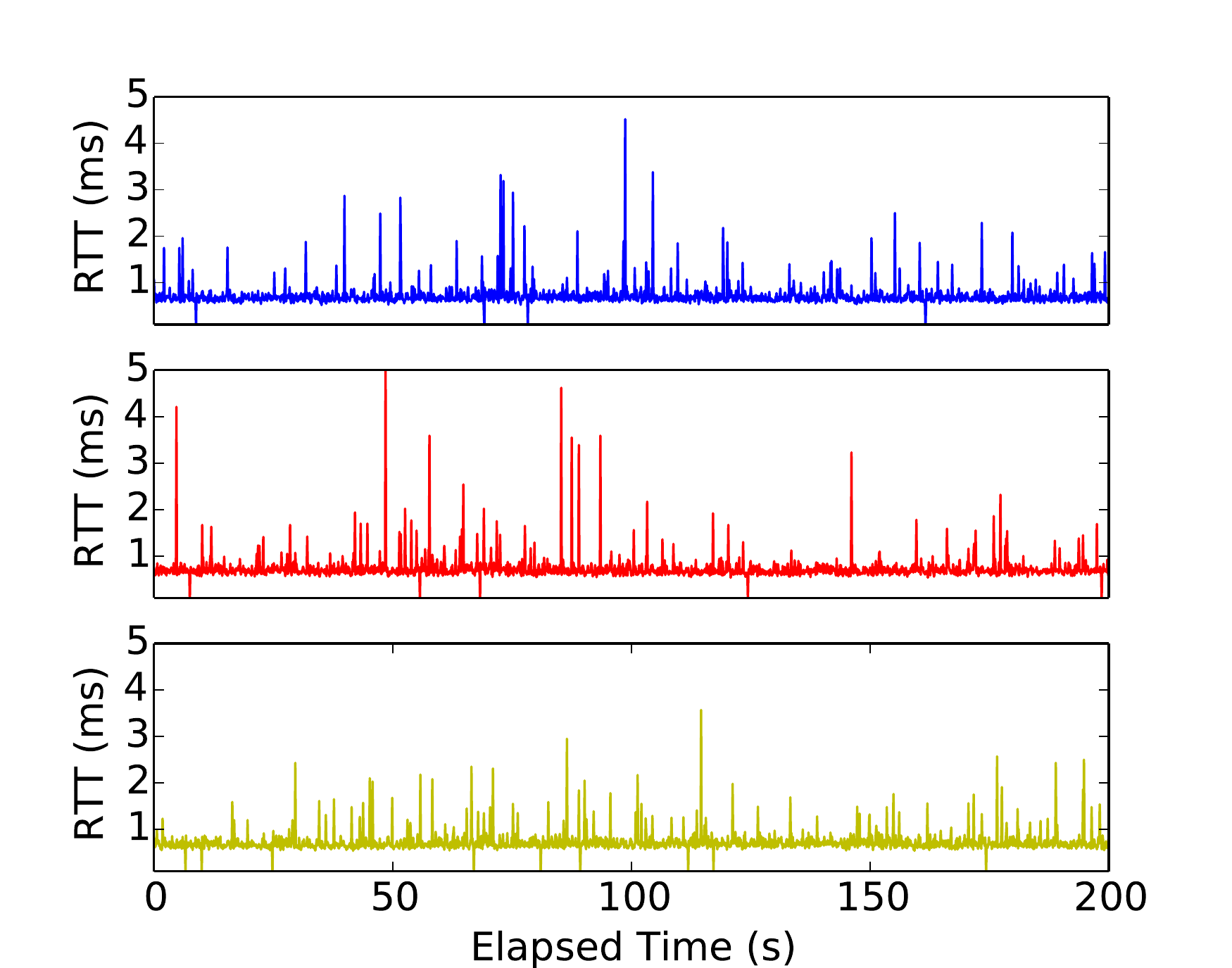}
	\vspace{-0.0cm} \caption{RTT of three paths between two pods of a fat-tree in Mininet.} %
	\label{fig:RTTfluctuation} \vspace{-0.3cm}
\end{figure}

We experimentally validate this observation using Mininet \cite{HHJL12} with real traffic traces from data centers. We construct a 6-pod unsubscribed fat-tree, with 3 hosts per rack. Traffic traces from a web search cluster \cite{AGMP10} are used to generate flows with average link load of 0.3, and we configure 3 hosts in one rack to {ping} 3 hosts of another rack in a different pod, respectively. A POX controller is configured to route the 3 ICMP sequences to 3 {distinct} paths between the two ToR switches. The interval of {ping} is 100~ms and the measurement lasts for 200 seconds. The RTT results are shown in Fig.~\ref{fig:RTTfluctuation}. It highlights two key characteristics of data center traffic: (1) RTT on a single path is low most of the time, indicating no congestion; and (2) flash congestion, which results in occasional peaks in the RTT, occurs \emph{independently} on different paths --- it is rare that all paths are congested simultaneously.

This form of path diversity motivates the idea of flow replication \cite{XL14}.
By trading a small amount of traffic redundancy for a higher degree of connectivity, replication considerably lowers the probability of transmission experiencing long latency. Theoretically speaking, if the proportion of congested paths between two end hosts is $p$, then the probability of a flow being delayed due to congestion is lowered from $p$ to $p^2$ after replication. Since the hot spots in data center networks are typically scarce, we have $p \ll 1$, such that $p^2 \ll p$.

\subsection{Testbed Verification}

\begin{figure*}[ht!]
	\centering
	\begin{subfigure}[t]{0.38\textwidth}
		\centering
		\includegraphics[width=\textwidth]{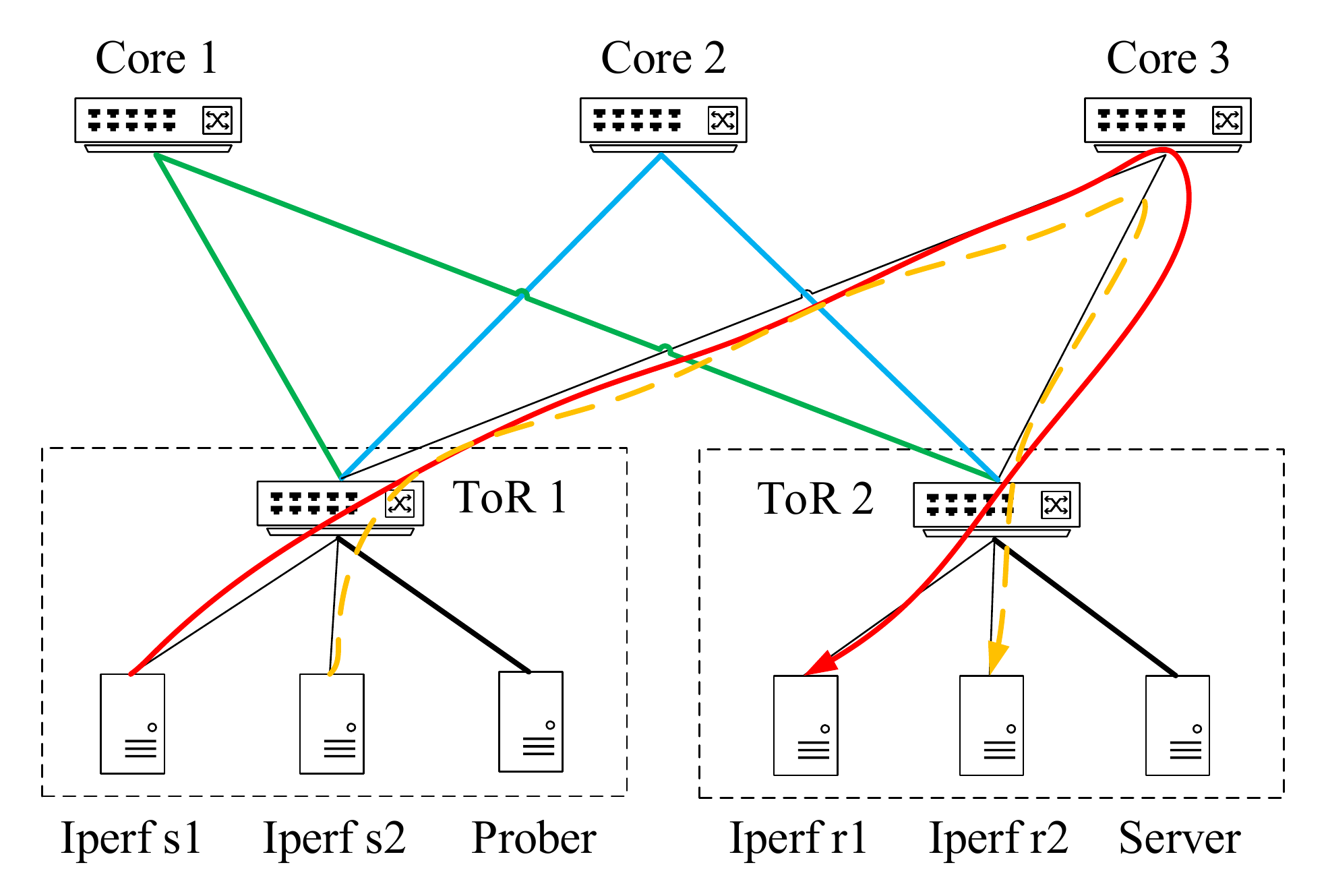}
		\caption{Experiment leaf-spine topology.}
	\end{subfigure}%
	\begin{subfigure}[t]{0.32\textwidth}
		\centering
		\includegraphics[width=\textwidth]{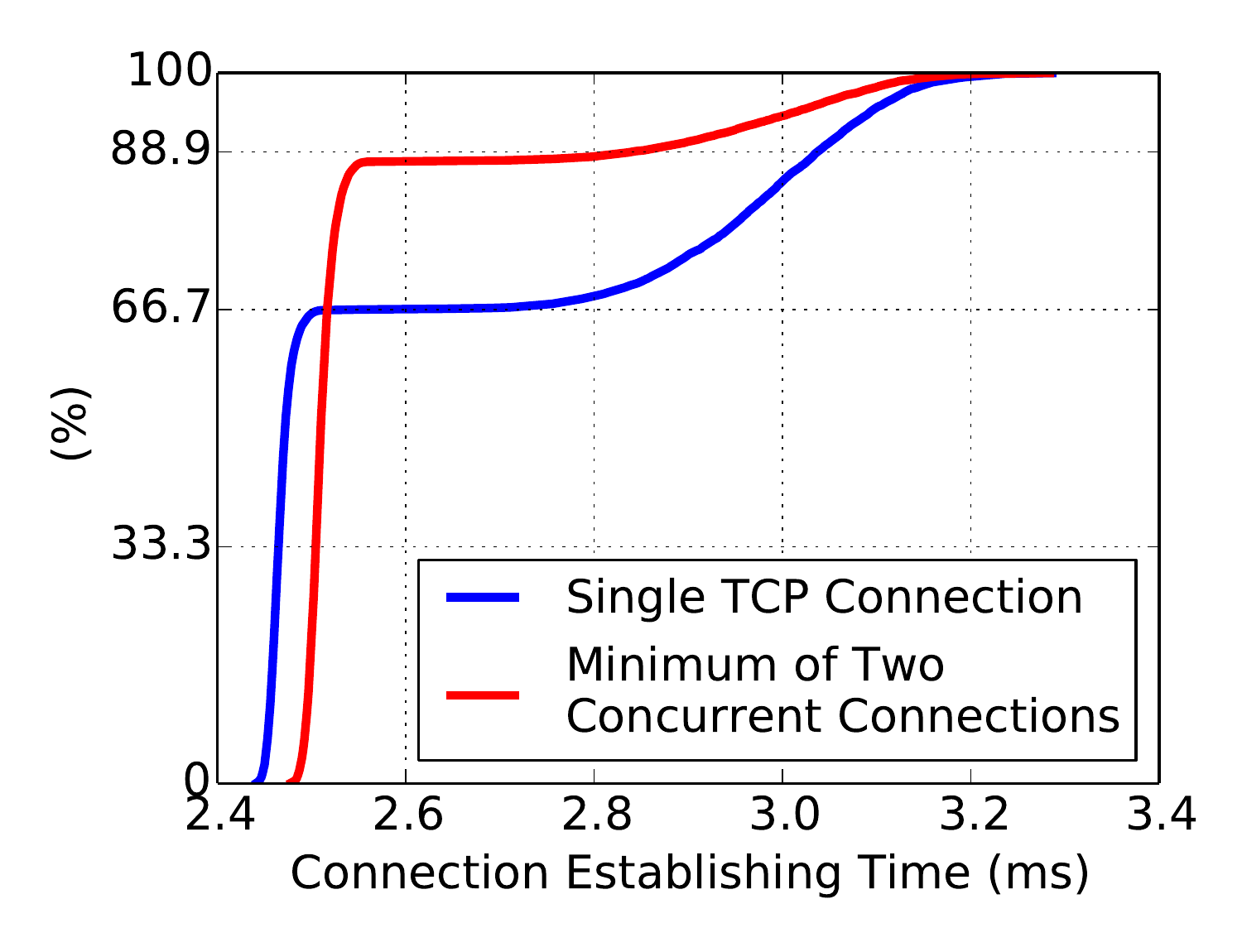}
		\caption{CDF of the measured RTTs.}
	\end{subfigure}%
	\begin{subfigure}[t]{0.32\textwidth}
		\centering
		\includegraphics[width=\textwidth]{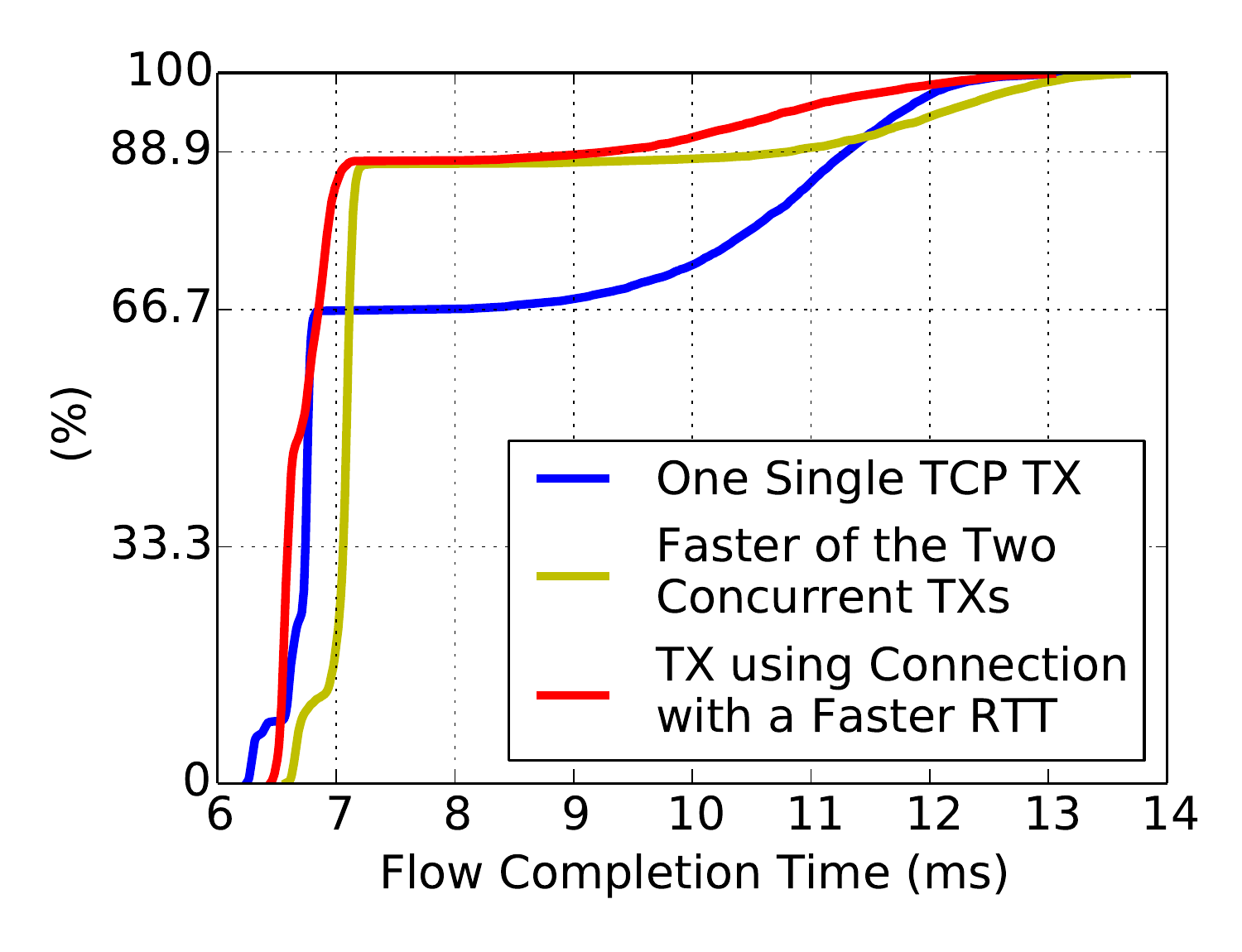}
		\caption{CDF of 100KB mice flow FCTs.}
	\end{subfigure}
	\vspace{-0.0cm}\caption{Experimental evaluation results to verify our motivation for flow replication.} %
	\label{fig:motivation_evaluation}\vspace{-0.3cm}
\end{figure*}

The above intuition is verified in our testbed. We establish a small leaf-spine topology with 3 paths available between two racks as shown in Fig.~\ref{fig:motivation_evaluation}(a). More detail about the testbed setup will be explained in Sec.~\ref{sec:testbed_setup}. 

We generate long-live flows using {Iperf} that congest one of the three paths as illustrated in Fig.~\ref{fig:motivation_evaluation}(a). Two {Iperf} senders, s1 and s2 in the left rack, are communicating with r1 and r2 in the right rack, respectively. We are able to confirm that two {Iperf} flows are routed to the same path and they are sending at half the link rate ($\sim$500Mbps) each. Meanwhile, the other two paths are idle.

We then measure RTT between the \emph{prober} in the left rack and the \emph{server} in the right rack, which is shown in Fig.~\ref{fig:motivation_evaluation}(b). The RTT is measured at the application layer during TCP handshaking. Specifically, the \emph{prober} opens a TCP connection by sending a SYN packet to the {\em server} and starts timing. The timing stops as soon as the connection is established successfully. We collect 10K RTT samples for each setting. As seen from Fig.~\ref{fig:motivation_evaluation}(b), the RTT distribution in a real testbed matches our probability analysis in the motivation example well. That is, with ECMP, a redundant TCP connection can lower the probability of choosing a congested path from $p$ ($\frac{1}{3}$ in this case) to $p^2$ ($\frac{1}{9}$).

We also collect FCTs of 100~KB mice flows, whose CDFs are illustrated in Fig.~\ref{fig:motivation_evaluation}(c), using three methods: (1) Send the flow with one TCP. (2) Send the same flow using two concurrent TCP connections, and record the FCT of the first one that finishes. (3) Start two TCP connections at the same time first, then send the payload through the faster connection with a smaller RTT. Clearly, the CDFs in Fig.~\ref{fig:motivation_evaluation}(c) show a similar trend to the RTT distribution in Fig.~\ref{fig:motivation_evaluation}(b). Moreover, the RTT of probe packets can reasonably reflect the congestion of the chosen path, since methods (2) and (3) are similar in distribution. These observations motivate the idea of RepFlow and RepSYN.

\subsection{{RepNet} Design}
\label{sec:repnet_design}
{RepNet} comprises of two mechanisms: RepFlow \cite{XL14} and RepSYN. In both mechanisms, only mice flows less than 100~KB are replicated. This can be easily changed for different networks.

RepFlow realizes flow replication by simply creating two TCP sockets for transmitting identical data for the same flow, as proposed in our previous work \cite{XL14}. Though conceptually simple, RepFlow doubles the number of bytes to be transmitted for mice flows. Further, it may aggravate throughput collapse in incast scenarios, when multiple flows sending concurrently to the same destination host \cite{VPSK09}.

We thus design RepSYN to overcome RepFlow's shortcomings. The idea is simple: Before transmitting data, we establish two TCP connections as in RepFlow. However data is only transmitted using the first established connection, and the other is ended immediately. Essentially SYN is used to probe the network and find a better path. The delay experienced by the SYN reflects the latest congestion condition of the corresponding path. RepSYN only replicates SYN packets and clearly does not aggravate incast compared to TCP.

\section{Implementation}
\label{sec:implementation}

We now describe our implementation of {RepNet} with {\tt node}. The source code is available online \cite{repnet}. 


\subsection{Why {\tt node}?} 

On a high level, {\tt node} is a highly scalable platform for real-time server-side networked applications. It combines single-threaded, non-blocking socket with the even-driven philosophy of JavaScript. It runs on Google V8 engine with core libraries optimized for performance and scalability. For more details see \cite{node-website}. 

The first reason for choosing {\tt node} is \emph{efficiency}. Replication introduces the overhead of launching additional TCP connections. To provide maximal latency improvements, we need to minimize this overhead. This rules out a multi-threaded implementation using for example {\tt Tornado} or {\tt Thrift} \cite{SAK07}. For one thing, replicating mice flows nearly doubles the number of concurrent connections a server needs to handle. For the other, the necessary status synchronization between the original connection and its replica demands communication or shared memory across threads. For applications with I/O from a large number of concurrent connections, a multi-threaded RepFlow will be burdened by frequent thread switching and synchronization \cite{Tilkov-2010-Node} with poor performance and scalability. In fact, we tried to implement RepNet on {\tt Thrift} based on python, and found that the performance is unacceptable. 

{\tt node} satisfies our requirement for high efficiency. Specifically, its non-blocking I/O model in a single thread greatly alleviates the CPU overhead. Asynchronous sockets in {\tt node} also avoid the expensive synchronization between the two connections of RepFlow. For example, it is complex to choose a quicker completion between two \texttt{socket.read} operations using blocking sockets: three threads and their status sharing will be needed. Instead, {\tt node} relies on callback of the \texttt{`data'} event to handle multiple connections in one thread, which greatly reduces complexity. The thread stack memory footprint (typically 2MB per thread) is also reduced.

The second reason we choose {\tt node} is that it is widely deployed in production systems for companies such as LinkedIn, Microsoft, etc. \cite{nodeatlinkedin}. Besides deployment in front-end web servers to handle user queries, a large number of companies and open source projects rely on \texttt{node} at the back-end for compatibility\footnote{\url{http://nodejs.org/industry/}}. \texttt{node} integrates smoothly with NoSQL data stores, e.g. MongoDB\footnote{Node.JS MongoDB Platform. \url{www.mongolab.com/node-js-platform}}, and caches, e.g. memcached\footnote{\url{https://nodejsmodules.org/pkg/memcached}}, and enables a full JavaScript stack for the ease of application development and maintenance. For these reasons, \texttt{node} is commonly used in data centers to fetch data across server machines.
Thus implementing {RepNet} on it is likely to benefit a large audience and generate immediate impact to the industry. 

\subsection{Overview}
\label{sec:imple-overview}

{RepNet} is based upon the \texttt{Net}\footnote{\url{http://nodejs.org/api/net.html}.} module, {\tt node}'s standard library for non-blocking socket programming. Similar to \texttt{Net}, {RepNet} exposes some socket functions, and wraps useful asynchronous network methods to create even-driven servers and clients, with additional low latency support by flow replication.

We implement {RepNet} with the following objectives:

{\bf Transparency.} {RepNet} should provide the same set of APIs as \texttt{Net}, making it transparent to applications. That is, to enable RepFlow, one only needs to include \texttt{require(`repnet')} instead of \texttt{require(`net')}, without changing anything else in the existing code.

{\bf Compatibility.} A {RepNet} server should be able to handle regular TCP connections at the same time. This is required as elephant flows are not replicated.

{RepNet} consists of two classes: {\tt RepNet.Socket} and {\tt RepNet.Server}. {\tt RepNet.Socket} implements a replication capable asynchronous socket at both ends of a connection. It maintains a single socket abstraction for applications while performing I/O over two TCP sockets. {\tt RepNet.Server} provides functions for listening for and managing both replicated and regular TCP connections. Note that {\tt RepNet.Server} does not have any application logic. Instead, it creates a connection listener at the server side, which responds to SYN packets by establishing a connection and emitting a connected {\tt RepNet.Socket} object in a corresponding callback for applications to use.

We now explain the high-level design and working of {RepNet} by examining the lifetime of a RepFlow transmission. The case of RepSYN is similar. First, the server runs a \texttt{RepNet.Server} that listens on two distinct ports. This is to make sure that the original and replicated flows have different five-tuples and traverse different paths with ECMP. When the client starts a RepFlow connection, a \texttt{RepNet.Socket} object is instantiated. Two \texttt{Net.Socket} objects, being two members of the \texttt{RepNet.Socket} object, will send SYN packets to the two ports on the receiver, respectively. They share the same source port number though, so the server can correctly recognize them among potentially many concurrent connections it has.

Now our server may not get the two SYN packets at the same time. To minimize delay, upon the arrival of the first SYN, the server responds immediately by emitting a new \texttt{RepNet.Socket}, using one member \texttt{Net.Socket} to process handshaking while creating another null \texttt{Net.Socket}. The first TCP connection is then established and ready for applications to use right away. 

The server now waits for the other connection. Its {\tt RepNet.Server} maintains a waiting list of connections --- represented by \texttt{<ip\_addr:port>} tuples --- whose replicas has yet to arrive. When the second SYN arrives, the server matches it against the waiting list, removes the connection from the list, and has the corresponding \texttt{RepNet.Socket} instantiate the other member \texttt{Net.Socket}. This second TCP connection will then proceed. At this point, both sides can send data using RepFlow, as two complete {\tt RepNet.Socket} objects. Note that the server also handles standard TCP connection. In this case a second SYN will never arrive and can be detected by timeout. 

Our implementation is based on {\tt node} 0.11.13. We introduce more details of our implementation in the following.

\subsection{Class: \texttt{RepNet.Socket}}
\label{sec:imple-repnetsocket}

\begin{figure}[hbt]
\vspace{-0.3cm}    
    \centering
    \includegraphics[width=0.4\textwidth]{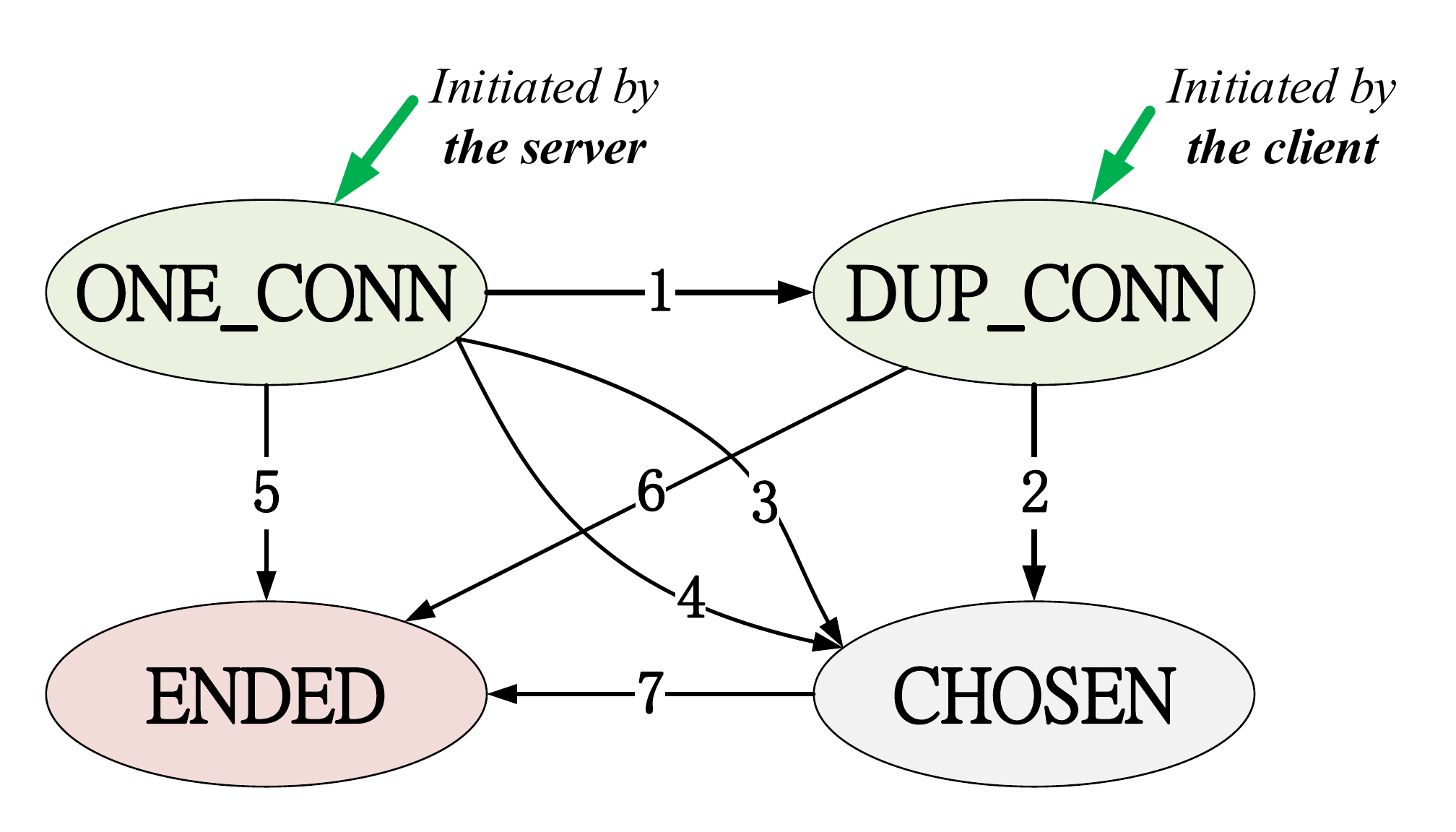}
    \vspace{-0.0cm} \caption{The FSM of \texttt{RepNet.Socket}.}
    \label{fig:fsm}
    \vspace{-0.4mm}
\end{figure}

The key difference between {\tt RepNet.Socket} and {\tt Net.Socket} is the I/O implementation. Since a {\tt RepNet.Socket} has two TCP sockets, a Finite State Machine (FSM) model is used to handle the asynchronous I/O across them. For brevity, all four states of the FSM are listed in Table~\ref{tab:statedescription}. Figure~\ref{fig:fsm} shows the possible state transitions with more explanation in Table~\ref{tab:statetrigger}. 

The client, who initiates the connection, always starts in {\tt DUP\_CONN}, and \texttt{socket.write()} in {RepNet} is done by calling \texttt{socket.write()} of both member \texttt{Net.Socket} objects to send data out. The server always starts in {\tt ONE\_CONN} waiting for the other SYN to arrive, and when it does enters {\tt DUP\_CONN}. In both states read operations are handled in the callback of a \texttt{`data'} event. A counter is added for each connection to coordinate the detection of new data. As soon as new chunks of buffer are received, \texttt{RepNet.Socket} emits its \texttt{`data'} event to the application.

For the server, if there are writes in {\tt ONE\_CONN}, they are performed on the active connection immediately and archived for the other connection with the associated data. If the archived data exceeds a threshold, the server enters {\tt CHOSEN} and disregards the other connection. The server may also enter {\tt CHOSEN} after timeout on waiting for the other connection, which corresponds to the standard TCP. 

\begin{table*}[hbt]
\small
\centering
\begin{tabular}{c|p{8.1cm}|c|p{3.5cm}}
\hline
\textbf{State}   & \multicolumn{1}{c|}{\textbf{Description}}                                                                                                                                                                          & \textbf{On Waiting List} & \multicolumn{1}{c}{\textbf{Performing I/O on}} \\ \hline
{\tt ONE\_CONN} & Only one {\tt Net.Socket} is open. The other one is pending.                                                                           & Yes & The only connection.                           \\ \hline
{\tt DUP\_CONN}  & Both member {\tt Net.Socket} objects are open.                                                                                                    & No & Both connections.                              \\ \hline
{\tt CHOSEN}     & One of {\tt Net.Socket} objects is no longer valid.        & Depend on State & The chosen connection.                  \\ \hline
{\tt ENDED}      & The {\tt RepNet.Socket} is ended.                                                                                                                                                                      & No & N/A                                            \\ \hline
\end{tabular}
\vspace{-0mm}
\caption{All states in the FSM.}%
\label{tab:statedescription}
\vspace{-0mm}
\end{table*}

\begin{table*}[hbt]
\centering
\small
\begin{tabular}{c|p{6.6cm}|p{8cm}}
\hline
\textbf{Transition} & \multicolumn{1}{c|}{\textbf{Trigger}}                                            & \multicolumn{1}{c}{\textbf{Additional Consequence}}                                                                                            \\ \hline
1                   & The slower connection is detected at the server.       & The corresponding flow is removed from the waiting list. The replicated connection is binded with the matching one. \\ \hline
2                   & One connection raises an exception, or emits an \texttt{`error'} event.       & The abnormal connection is abandoned by calling the \texttt{destroy()} function and resetting the other end.                                   \\ \hline
3                   & The corresponding flow in the waiting list is timed out. & The item is deleted from the waiting list.                                                                                    \\ \hline
4                   & The archived data for writes exceeds the threshold.          & The corresponding flow will NOT be removed from the  waiting list until the second SYN arrives for correctness.                      \\ \hline
5, 6, 7             & Both connections are destroyed or ended.                                         &                                                                                                                                                \\ \hline
\end{tabular}
\vspace{-0mm}
\caption{Trigger of the state transitions}%
\label{tab:statetrigger}
\vspace{-0mm}
\end{table*}

\subsection{Class: \texttt{RepNet.Server}}

\texttt{RepNet.Server} has two \texttt{Net.Server} objects which listen on two distinct ports. The key component we add is the waiting list for RepFlow which we explain now.

The waiting list is a frequently updated queue. Each flow in the waiting list has three fields: \texttt{TTL}, \texttt{flowID} (the client's \texttt{<ip\_addr:port>} tuple), and \texttt{handle} (a pointer to the corresponding \texttt{RepNet.Socket} instance). 

There are three ways to update the list:

{\bf Push.} If a new SYN arrives and finds no match in the list, a new \texttt{RepNet.Socket} object is emitted and its corresponding flow will be pushed to the list.

{\bf Delete.} If a new SYN arrives and it matches with an existing flow, the corresponding \texttt{RepNet.Socket} object is then completed and this flow is removed from the list.

{\bf Timeout.} If the flow stays on the list for too long to be matched, it is timed out and removed. This timeout can be adjusted by setting the \texttt{WL\_TIMEOUT} option. The default is equal to $RTO$ of the network. A higher value of \texttt{WL\_TIMEOUT} may decrease the probability of matching failures, at the cost of increasing computation and memory.


Note that to achieve transparency by exposing the same APIs as \texttt{Net.Server}, the constructor of \texttt{RepNet.Server} accepts only one port number parameter. It simply advances the number by one for the second port. An \texttt{error} event will be emitted if either of the port is already in use.

\subsection{RepSYN to Alleviate Incast}


As explained in Sec.~\ref{sec:repnet_design}, we propose RepSYN to alleviate RepFlow's drawbacks in incast scenarios. 
A RepSYN client can work compatibly with a \texttt{RepNet.Server}. Specifically, once the second connection is established and the server-side socket enters {\tt DUP\_CONN}, it would be reset immediately by the client to trigger the transition to {\tt CHOSEN} in Table~\ref{tab:statetrigger}. RepSYN can be activated by setting the \texttt{Flag\_RepSYN} flag of the \texttt{RepNet.Socket} object. 





\section{Testbed Evaluation}
\label{sec:evaluation}

We present our testbed evaluation of RepNet in this section.

\subsection{Testbed Setup}
\label{sec:testbed_setup}
Our testbed uses Pronto 3295 48-port Gigabit Ethernet switches with 4MB shared buffer. The switch OS is PicOS 2.04 with ECMP enabled. Each server has an Intel E5-1410 2.8GHz CPU (8-thread quad-core), 8GB memory, and a Broadcom BCM5719 NetXtreme Gigabit Ethernet NIC.

The servers run Debian 6.0 64-bit Linux, kernel version 2.6.38.3. We change $\text{RTO}_{\text{min}}$ to 10~ms in order to remedy the impact of incast and packet retransmission \cite{VPSK09}. We found that setting it to a value lower than 10~ms leads to system instability in our testbed. The initial window size is 3, i.e. about 4.5~KB payload. The initial RTO is 3 seconds by default in the kernel, which influences our experiments in cases where TCP connections fail to establish at the first time. We tried to set it to a smaller value, but found that kernel panics occur frequently because of fatal errors experienced by the TCP keep-alive timer.

\begin{figure}[hbt]
    \vspace{-0.0cm} \centering
    \includegraphics[width=0.35\textwidth]{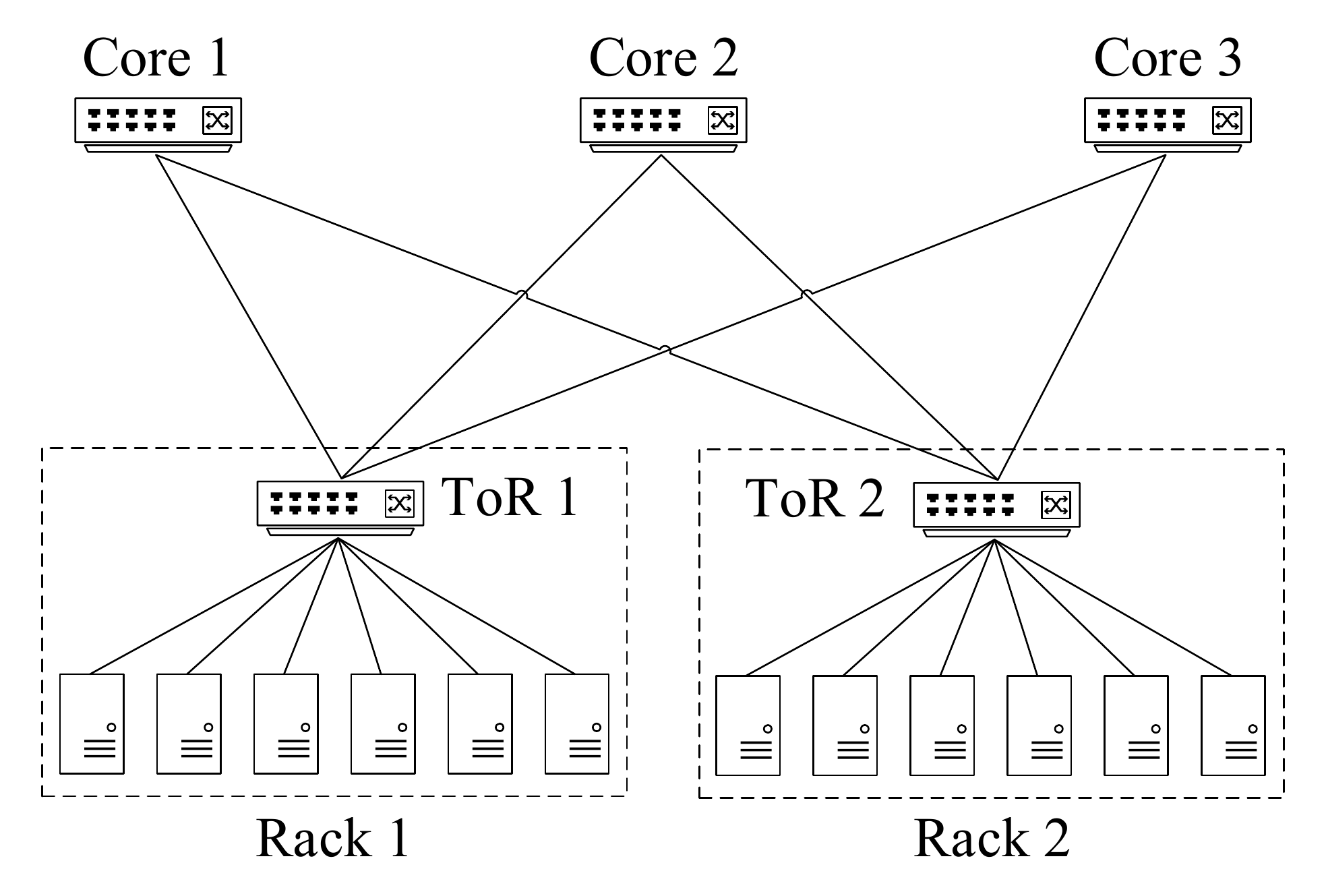}
    \vspace{-0cm} \caption{The leaf-spine topology of the testbed.} %
    \label{fig:testbed-topo} \vspace{-0.3cm}
\end{figure}

{\bf Topology.}
The testbed uses a leaf-spine topology as depicted in Fig.~\ref{fig:testbed-topo} which is widely used in production data centers \cite{AYSK13}. There are 12 servers organized in 2 racks, and 3 spine switches which provide up to 3 equal-cost paths between two hosts under different ToRs. The {ping} RTT is $\sim$178~$\mu s$ across racks.
The topology is oversubscribed at 2:1 when all hosts are used. We also conduct experiments without oversubscription, by shutting down half of the servers in each rack.

\subsection{Empirical Traffic Performance}%
\label{sec:testbed-empirical}

\begin{figure*}[t!]
    \centering
    \begin{subfigure}[t]{0.32\textwidth}
        \centering
        \includegraphics[width=\textwidth]{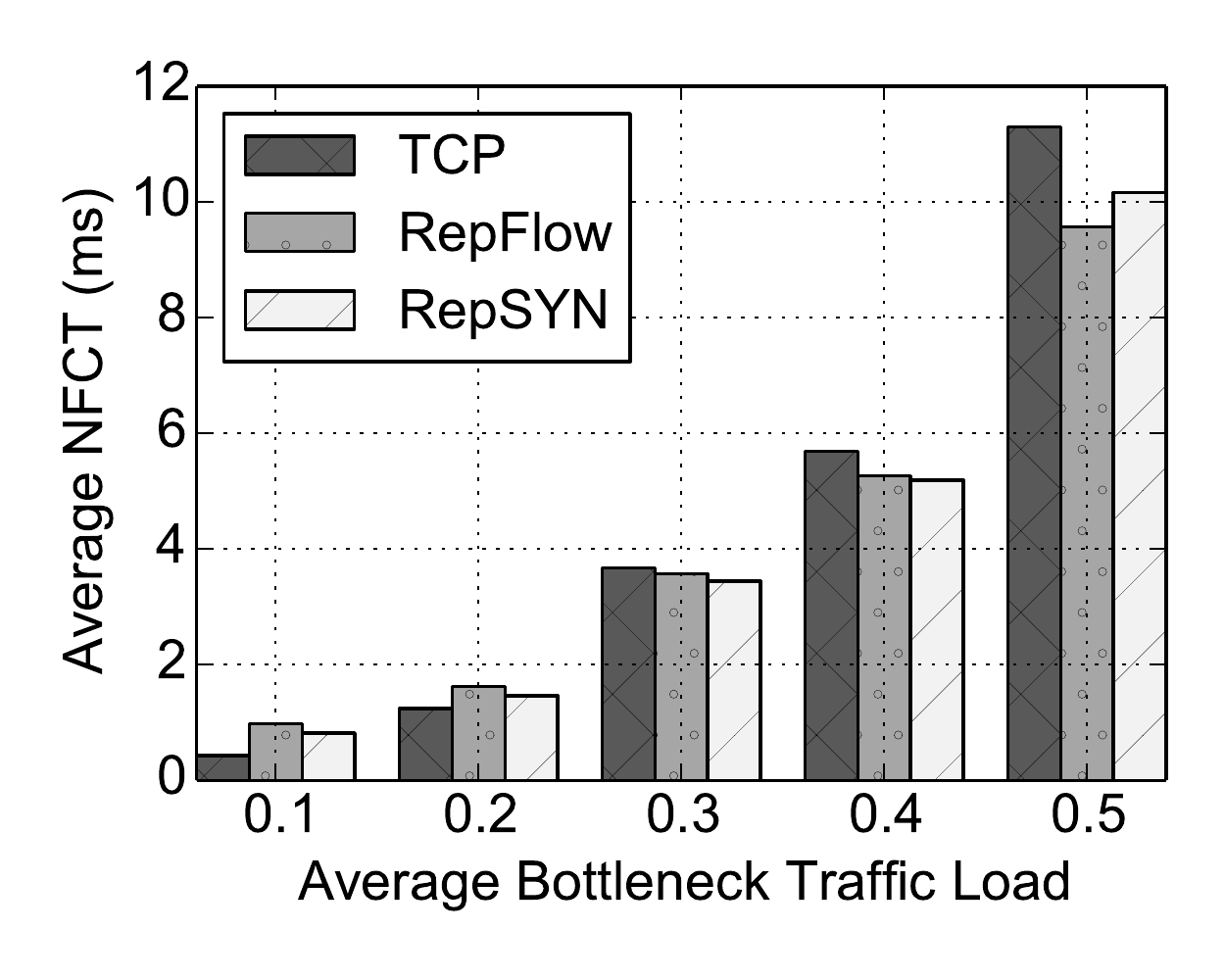}
    \end{subfigure}%
    ~ 
    \begin{subfigure}[t]{0.32\textwidth}
        \centering
        \includegraphics[width=\textwidth]{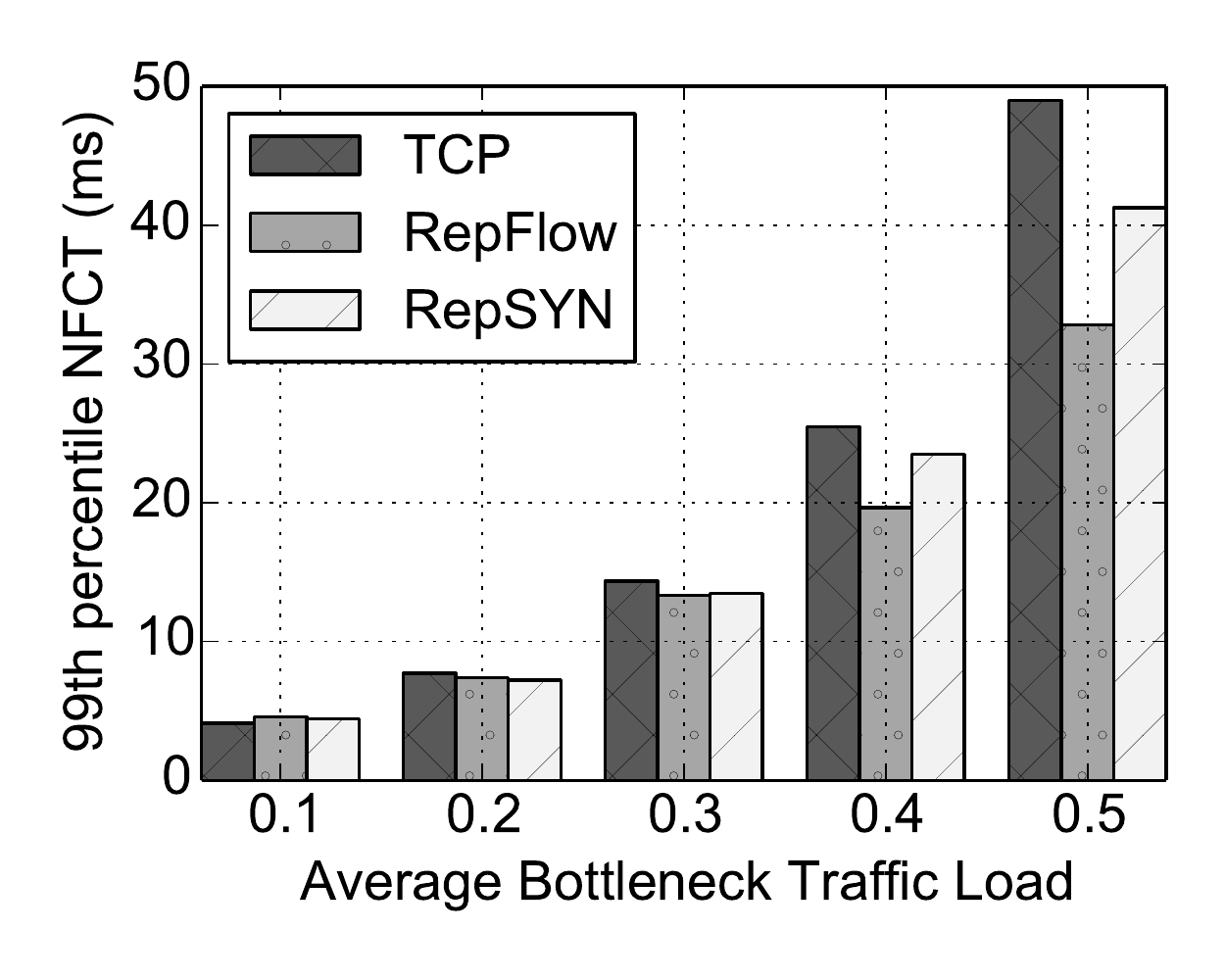}
    \end{subfigure}%
    ~ 
    \begin{subfigure}[t]{0.32\textwidth}
        \centering
        \includegraphics[width=\textwidth]{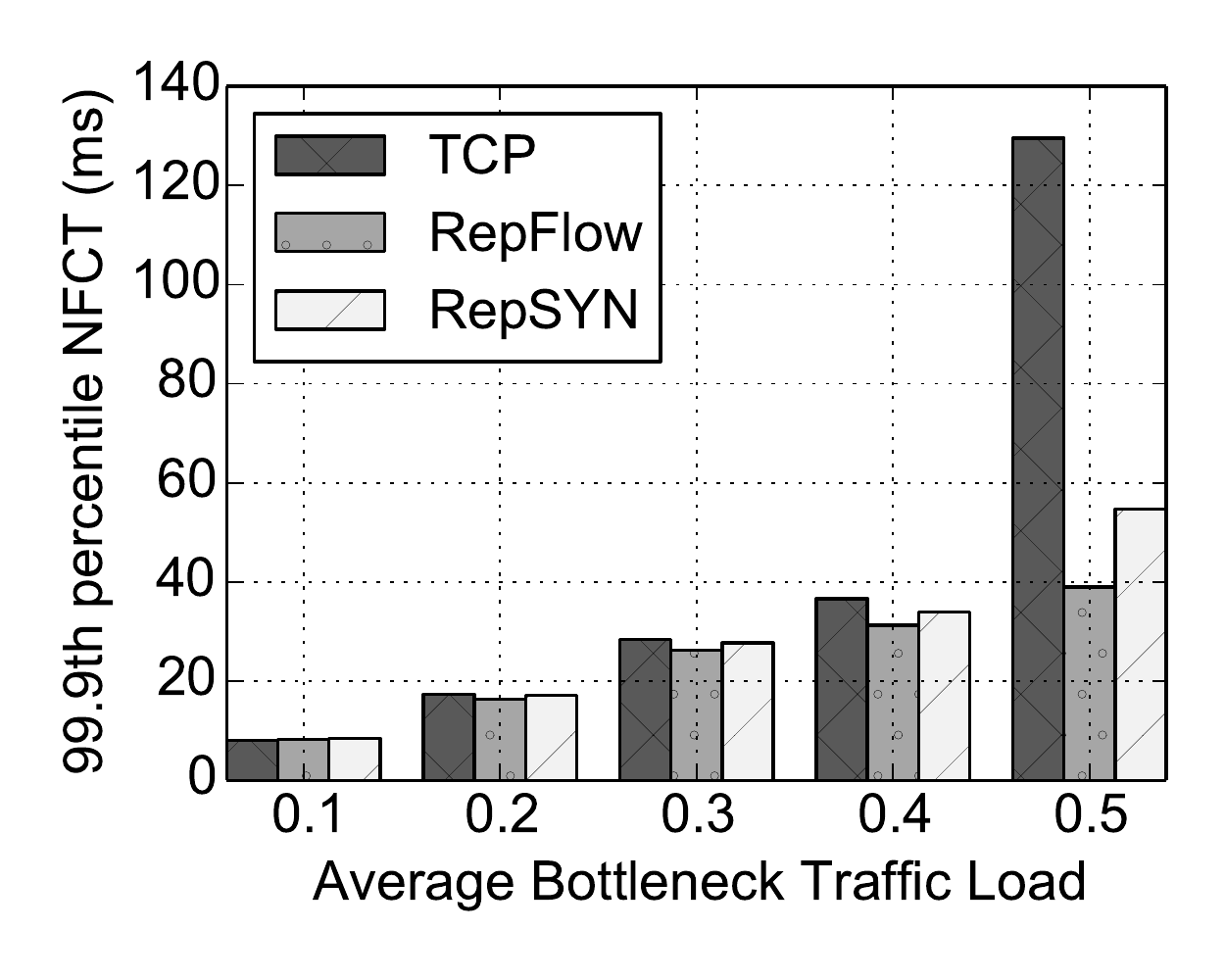}
    \end{subfigure}
    \vspace{-0.0cm}\caption{NFCT comparison when network oversubscription is 1:1.} %
    \label{fig:1NFCT}\vspace{-0.0cm}
\end{figure*}

\begin{figure*}[ht!]
    \centering
    \begin{subfigure}[t]{0.32\textwidth}
        \centering
        \includegraphics[width=\textwidth]{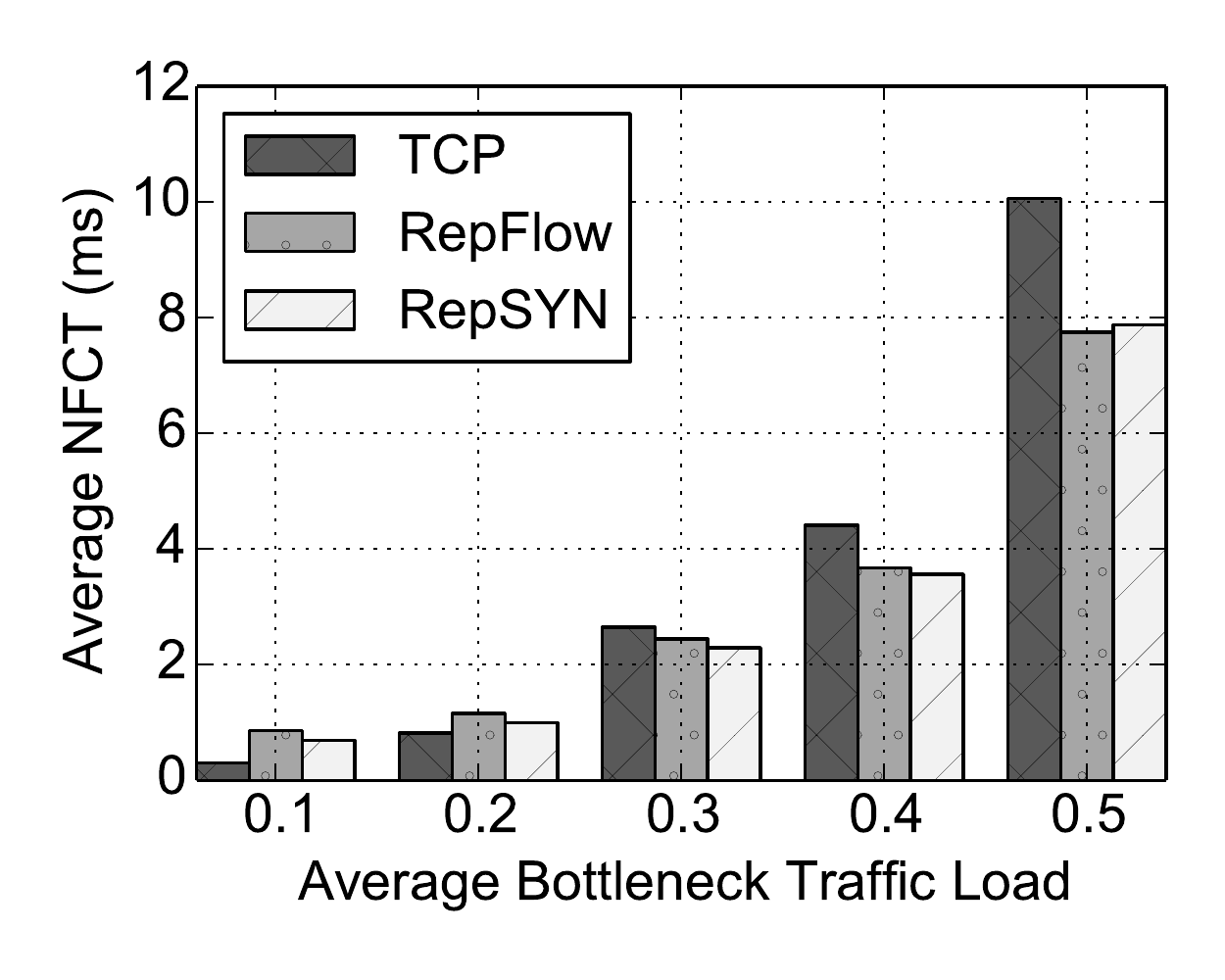}
    \end{subfigure}%
    ~ 
    \begin{subfigure}[t]{0.32\textwidth}
        \centering
        \includegraphics[width=\textwidth]{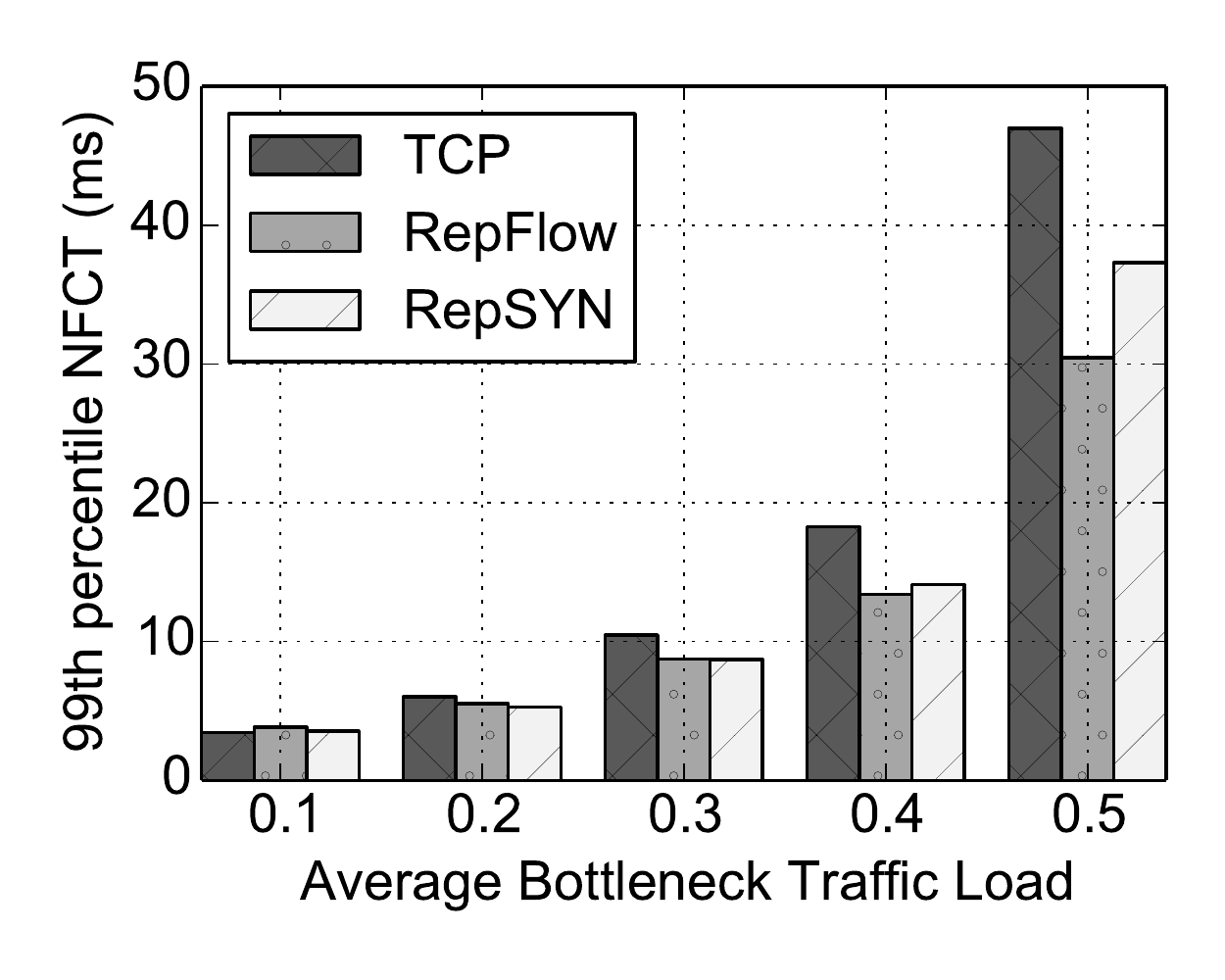}
    \end{subfigure}%
    ~ 
    \begin{subfigure}[t]{0.32\textwidth}
        \centering
        \includegraphics[width=\textwidth]{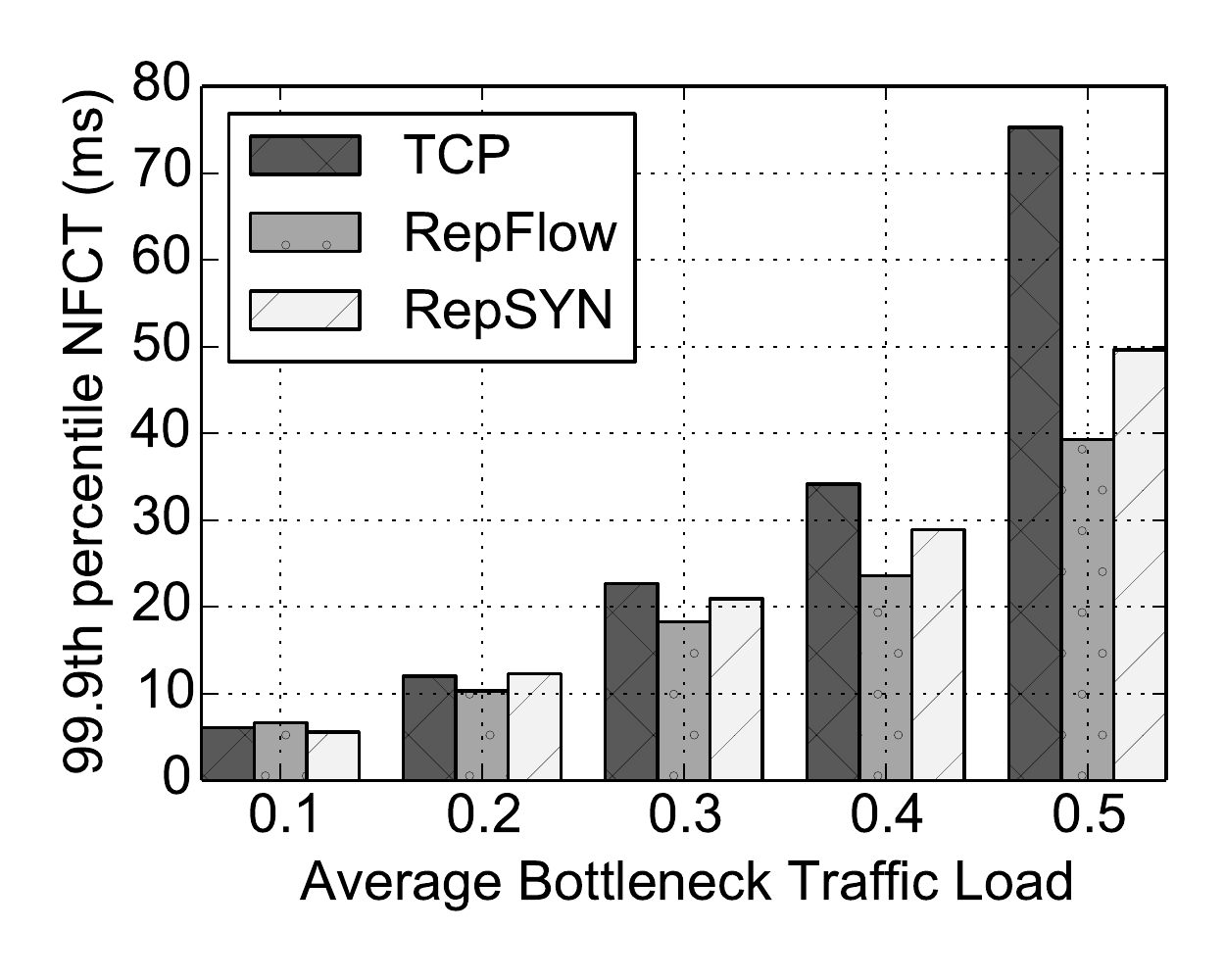}
    \end{subfigure}
    \vspace{-0.0cm} \caption{NFCT comparison when network oversubscription is 2:1.}%
    \label{fig:2NFCT} \vspace{-0.0cm}
\end{figure*}

{\bf Empirical Flow Size.}
We use the flow size distribution from a web search workload \cite{AGMP10} to drive our experiments. Most flows ($\sim$60\%) in this workload are mice flows smaller than 100KB, though over 95\% of the bytes are from 30\% of flows larger than 1MB.

Flows are generated between random pairs of servers in different racks following a Poisson process, with bottleneck traffic load varying from 0.1 to 0.5 for both the oversubscribed and non-oversubscribed settings. We notice that when the bottleneck load is higher than 0.5, packet drops and retransmissions become too frequent to conduct meaningful experiments. 
At each run, we collect and analyze flow size and completion time information from at least 200,000 flows for each scheme, and each experiment lasts for at least 6 machine hours.

{\bf Performance Metrics.}
We compare RepFlow and RepSYN against standard Linux TCP Cubic.
We use \emph{Normalized Flow Completion Time} (NFCT), defined as the measured FCT minus the kernel networking overhead for TCP as the performance metric. Kernel overhead includes for example socket creation, binding, context switching, etc., and varies depending on the OS and the networking stack. It is also possible to almost completely avoid this overhead using kernel bypass and other techniques \cite{KPTV12}. Thus we remove its impact in NFCT. Note that RepFlow and RepSYN incur more kernel overhead than TCP, which is included in their NFCT statistics by definition. We measure the kernel overhead of TCP as the average FCT of 100K flows of 1KB sent to localhost using our implementation without network latency, which is 6.82ms. More discussion on overhead is deferred to Sec.~\ref{sec:latency_overhead}.

Note that both RepFlow and RepSYN are completely working in the application layer, whose functionality is completely orthogonal to lower layer schemes. Thus we do not compare against these schemes.

\subsubsection{NFCT of Mice Flows}

First, we study the NFCT of mice flows. We compare three statistics, the average, the 99th percentile and 99.9th percentile NFCTs, to show RepFlow and RepSYN's impact on both the average and tail latency.

Fig.~\ref{fig:1NFCT} shows the results without oversubscription in the network.
Neither RepFlow nor RepSYN makes much difference when the load is low ($\leq 0.2$). As the load increases, RepFlow and RepSYN yield greater benefits in both average and tail latency. When the load is 0.5, RepFlow provides 15.3\%, 33.0\% and even 69.9\% reduction in average, 99th percentile, and 99.9th percentile NFCT, respectively. RepSYN also achieves 10.0\%, 15.8\% and 57.8\% reduction in average, 99th percentile, and 99.9th percentile NFCT, compared with TCP. 

An interesting observation is that when the load is high, RepFlow achieves significantly lower tail latency, while RepSYN becomes less beneficial. This is because compared to RepSYN, RepFlow with duplicated transmissions has a lower probability of experiencing packet losses which constitutes a great deal in tail latency.

When the network is oversubscribed at 2:1, the results are similar in general as shown in Fig.~\ref{fig:2NFCT}. RepFlow and RepSYN are in fact more beneficial in this case, because bursty traffic is more likely to appear at the second or third hop now, which can be avoided by choosing another available path. Therefore, in a production data center network where the topology is typically oversubscribed with many paths available, RepFlow and RepSYN are able to greatly reduce the tail latency and provide better performance.

We also study the impact of flow size on performance improvement. We divide all mice flows into 6 groups based on the minimum number of round trips needed to transmit by TCP. Fig.~\ref{fig:size-fct} illustrates the 99th percentile NFCT of these groups, when the load is 0.4. We can clearly see that RepFlow and RepSYN are equally beneficial for mice flows of different sizes. We observe the same result for different loads and oversubscription settings and omit the figures here.

\subsubsection{Incast}

We carefully study RepFlow and RepSYN's performance in incast scenrios here. In this experiment, whenever we generate a mice flow, we create another 10 flows of the same size with the same destination in parallel, resulting in a 11-to-1 incast pattern. For RepFlow it becomes 22-to-1 incast. Note the flow size distribution still flows the web search workload with both mice and elephants. 
\begin{figure}[t!]
\!\!\!\!\! \hspace{-0.2cm} \begin{minipage}[t]{0.235\textwidth}
    \centering
    \includegraphics[width=1.05\textwidth]{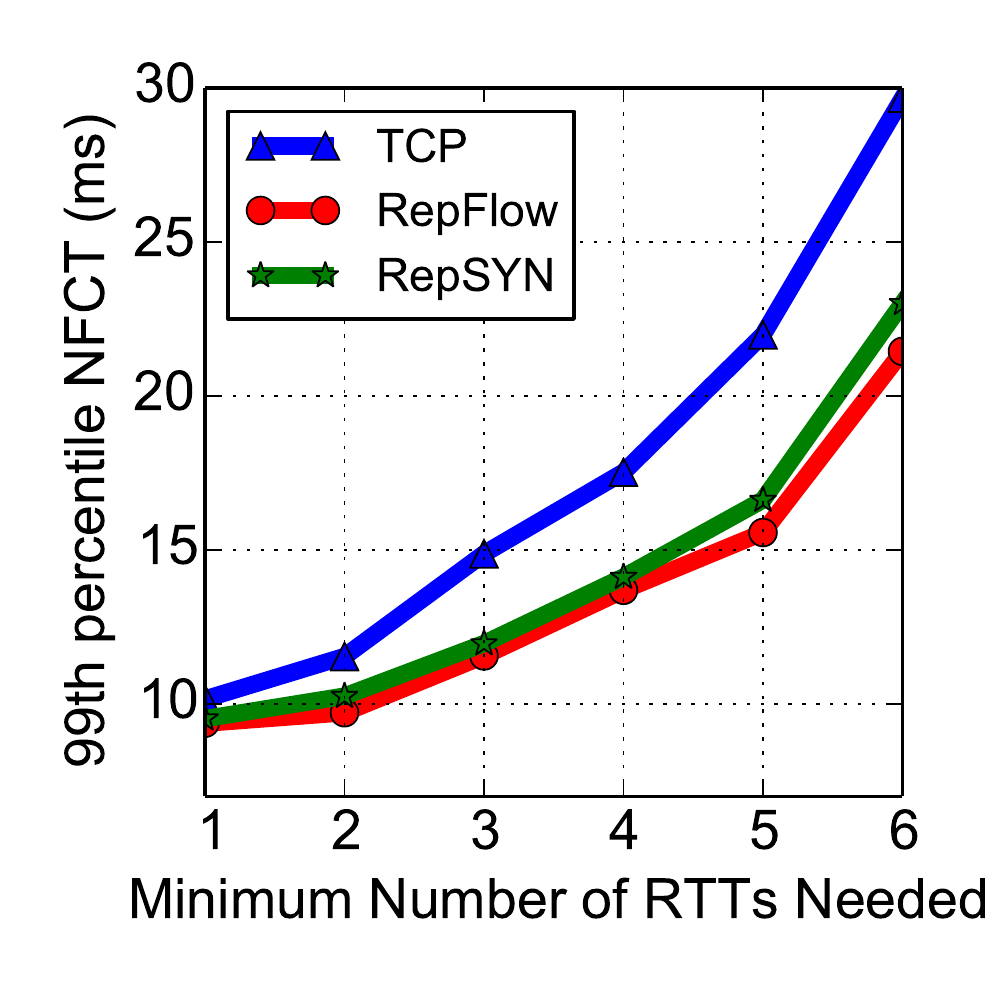}
    \vspace{-0.3cm} \caption{99th percentile NFCT comparison of flows with different sizes.} %
    \label{fig:size-fct}
\end{minipage} ~
\begin{minipage}[t]{0.235\textwidth}
    \centering
    \includegraphics[width=1.05\textwidth]{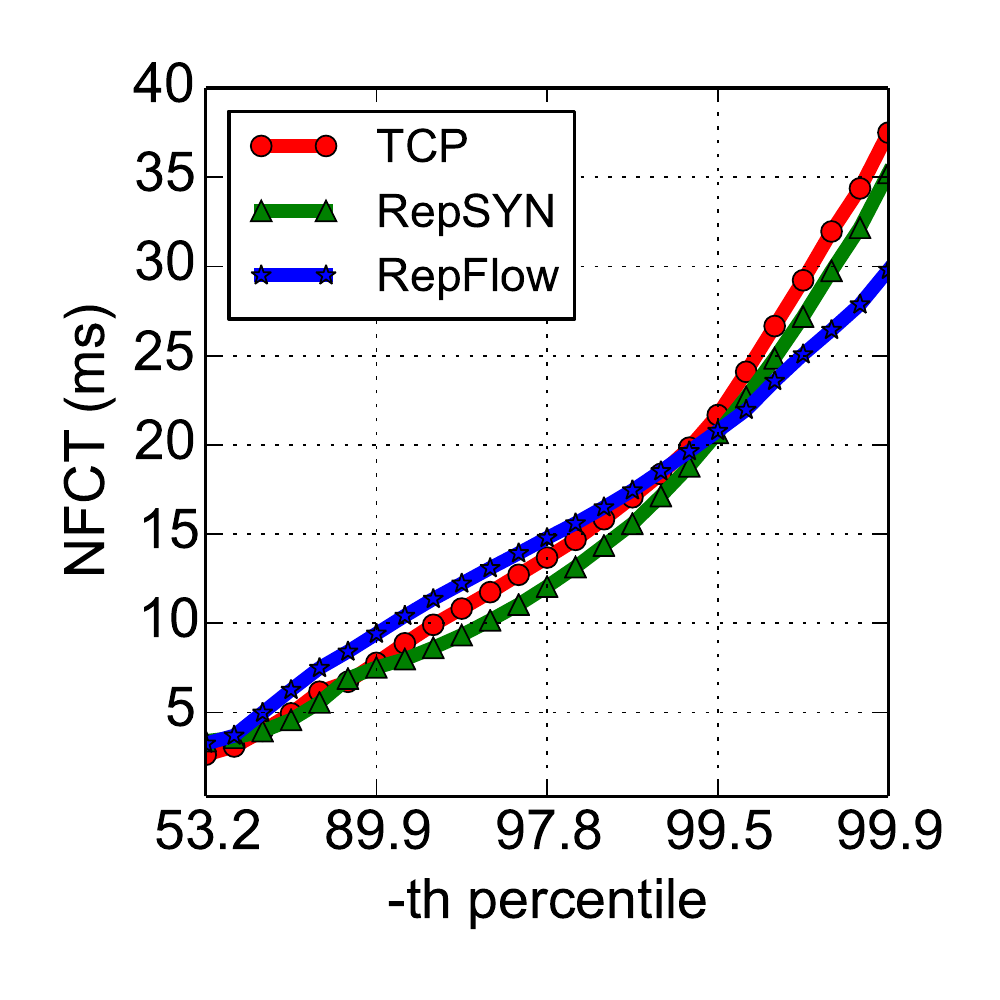}
    \vspace{-0.3cm} \caption{NFCT of mice flows in incast. Average bottleneck load is 0.2.}  \label{fig:incast-normal}
\end{minipage} 
\end{figure}

\begin{figure*}[t!]
\begin{minipage}[b]{0.64\linewidth}
    \begin{subfigure}[t]{0.5\textwidth}
        \centering
        \includegraphics[width=\textwidth]{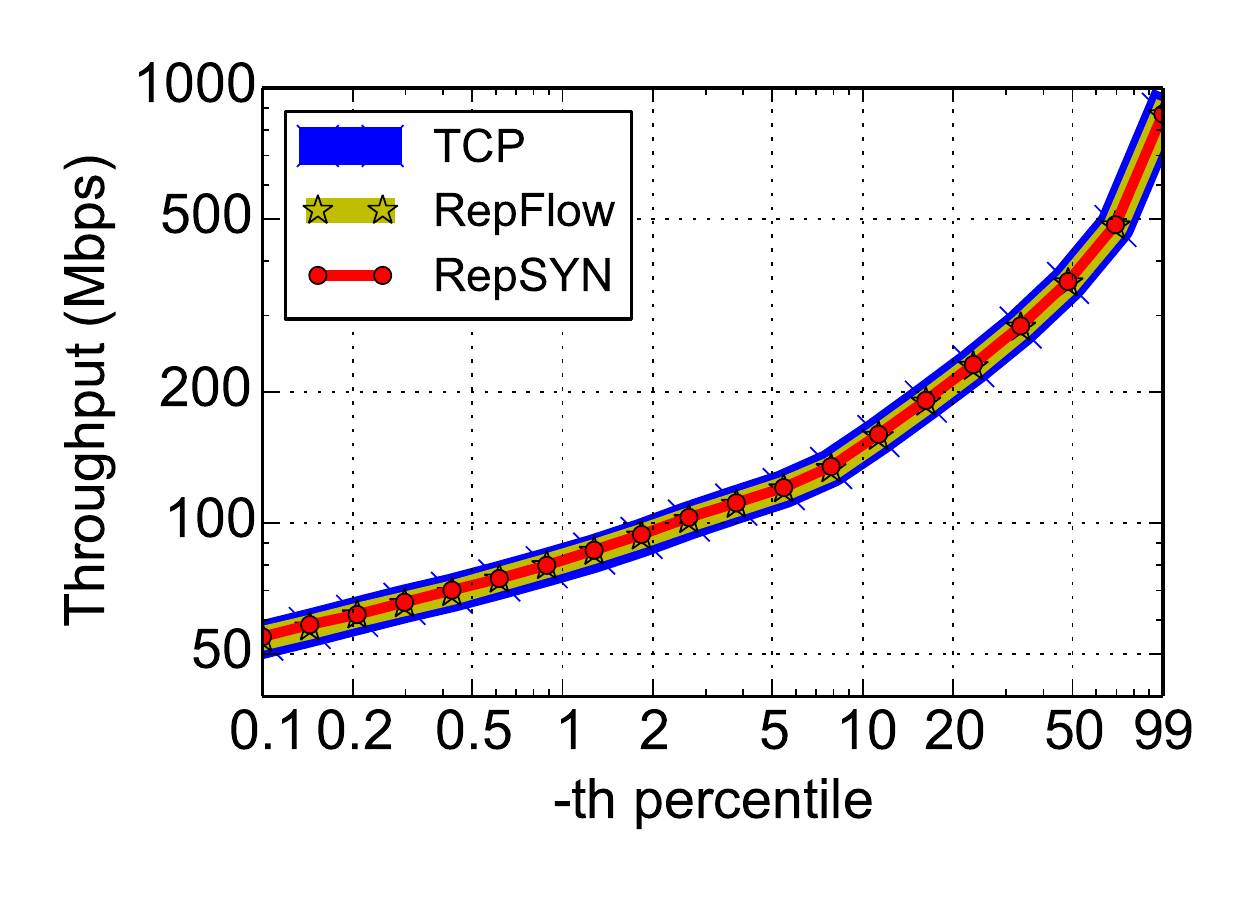}
        \vspace{-0.3cm} \caption{Low bottleneck traffic load of 0.2.} \label{fig:large-low}
    \end{subfigure}%
    \begin{subfigure}[t]{0.5\textwidth}
        \centering
        \includegraphics[width=\textwidth]{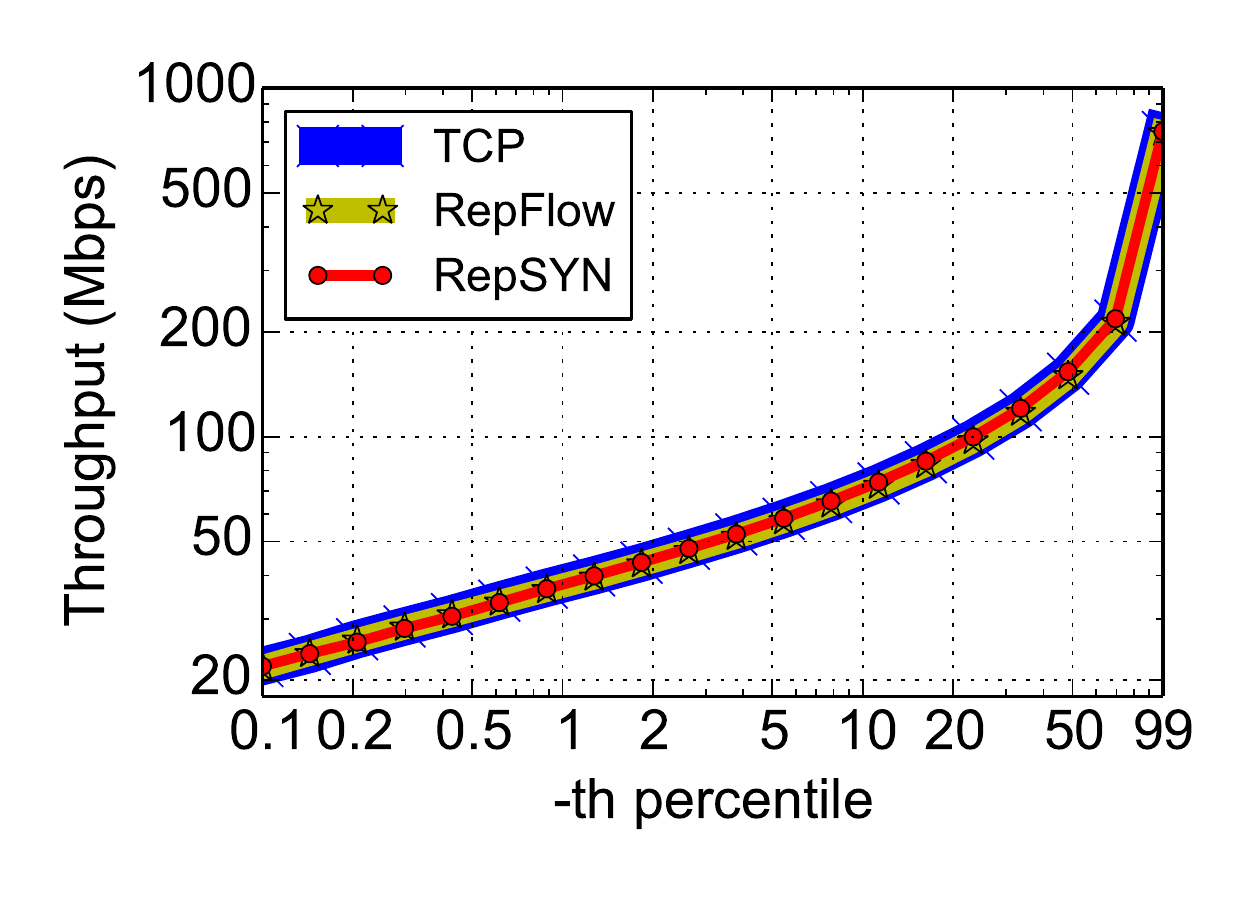}
        \vspace{-0.3cm} \caption{High bottleneck traffic load of 0.4.} \label{fig:large-high}
    \end{subfigure}%
    \vspace{-0.2cm} \caption{Throughput distribution of large flows.}
\end{minipage}
\quad
\begin{minipage}[b]{0.32\linewidth}
    \centering
    \includegraphics[width=\textwidth]{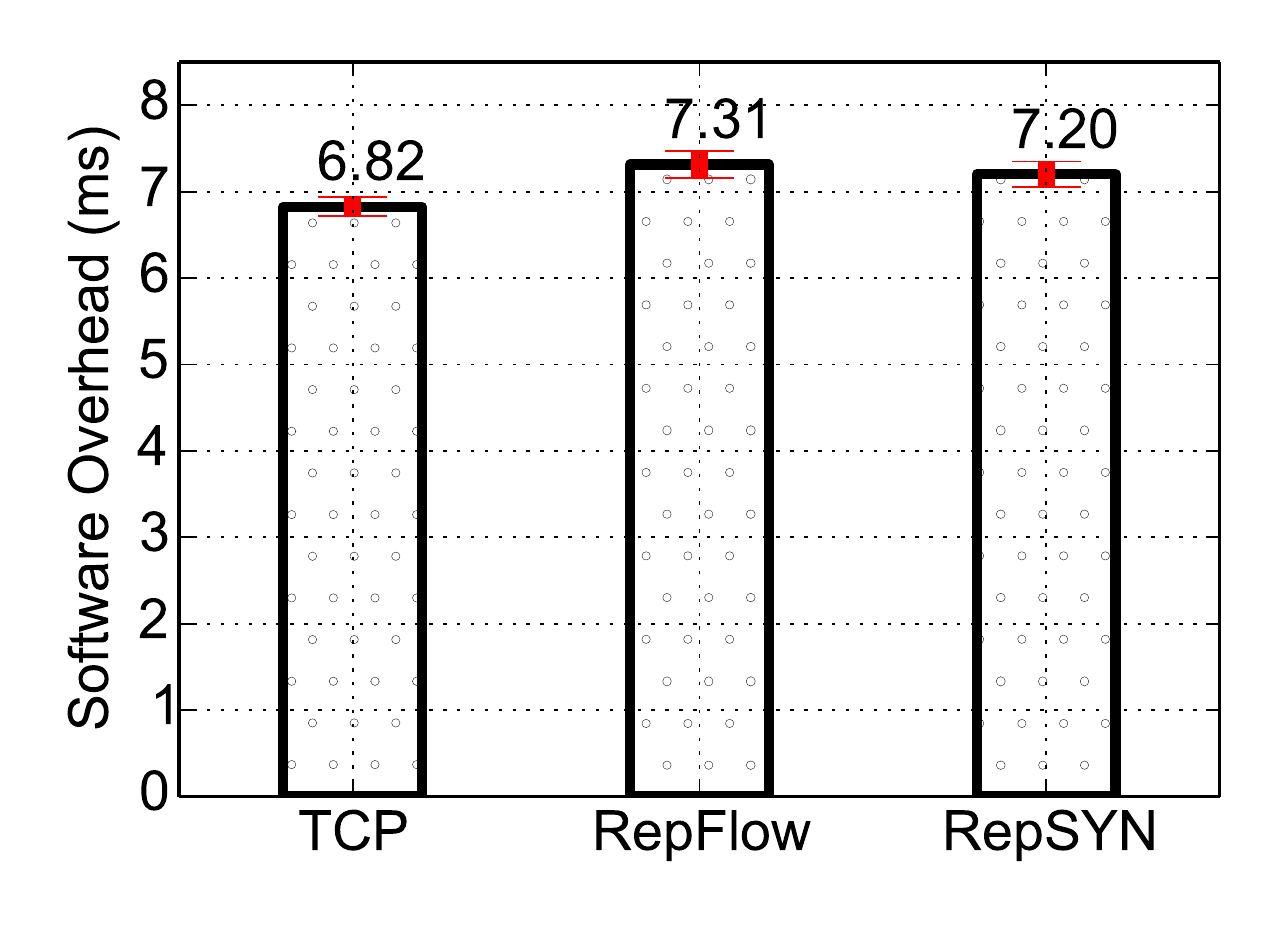}
    \vspace{-0.5cm} \caption{Kernel overhead comparison.} %
    \label{fig:overhead}
\end{minipage} \vspace{-0.3cm}
\end{figure*}

The performance is illustrated in Fig.~\ref{fig:incast-normal}. Note that the x-axis is in log scale, which shows more details about the tail latency. Though RepFlow is still able to cut the 99.9th percentile NFCT by 20.5\%, it is no longer beneficial in the 99th percentile, which is $\sim$400$\mu$s longer than TCP. Most flows experience longer delay using RepFlow. The benefit in the 99.9th percentile is because hash collision with elephants still contributes to the worst-case FCTs in our testbed. However, the benefit may be smaller if the concurrency of small flows was extremely high in incast. In those cases RepFlow could become a real burden.

Fig.~\ref{fig:incast-normal} shows that RepSYN, on the other hand, has 8.7\% and 6.0\% NFCT reductions in the 99th and 99.9th percentile, respectively. The slowest half of all flows are accelerated. Therefore, our suggestion for applications which incorporate serious many-to-one traffic patterns is to use RepSYN instead. Without aggravating the last hop congestion, RepSYN is still beneficial for reducing in-network latency.

\subsubsection{Impact on Large Flows}

Another possible concern is that RepFlow may degrade throughput for elephant flows due to the additional traffic it introduces. We plot throughput of elephants in both low and high loads in Fig.~\ref{fig:large-low} and Fig.~\ref{fig:large-high}, respectively. It is clear that throughput is not affected by RepFlow or RepSYN. The reason is simple: for data centers mice flows only account for a fraction of the total traffic \cite{AGMP10,GHJK09}, and replicating them thus cause little impact on elephants.

\subsubsection{Overhead of Replication}
\label{sec:latency_overhead}

We look at the additional kernel overhead of RepFlow and RepSYN due to the extra TCP connections and state management as in Sec.~\ref{sec:implementation}. We use the same method of obtaining kernel overhead of TCP --- measuring the FCT of 100K flows of 1KB sent to localhost --- for RepFlow and RepSYN. The result is shown in Fig.~\ref{fig:overhead} with error bars representing one standard deviation. Observe that on average, RepFlow incurs an extra 0.49ms of overhead, while RepSYN's overhead is only 0.32ms in our current implementation. Compared with tail NFCT which is more than 20ms, this overhead is negligible. Optimization such as kernel bypass \cite{KPTV12,latency_techniques} can further reduce this overhead though it is beyond the scope. 

\subsubsection{Discussion}
Finally, we comment that the testbed scale is small with limited multipath diversity. Both the kernel configuration and our implementation can be further optimized. Thus the results obtained shall be viewed as a conservative estimate of RepFlow and RepSYN's practical benefits in a production scale network with a large number of paths.

\subsection{Application-Level Performance}

Besides evaluation of flow-level performance with empirical traffic, a question remains unclear at this point: \emph{how much performance enhancement can we get by using {RepNet} for distributed applications in a cluster}? We answer this question by implementing a distributed bucket sort application in \texttt{node} in our testbed, and evaluating the job completion times with different transport mechanisms.

\subsubsection{A Sorting Application}

{\bf Application Design.}
We choose to implement bucket sort \cite{intro-algorithms-2001}, a classical distributed sorting algorithm, as an exemplar application with a partition-aggregation workflow. In bucket sort, there exists a master which coordinates the sorting process and several slave nodes which complete the sub-processes individually. Whenever the master has a large array of numbers (usually stored in a large text file) to sort, it divides the range of these values into a given number of non-overlapping groups, i.e. buckets. The master then scans the array, disseminates the elements to their corresponding buckets using TCP connections. Typically, each slave node holds a bucket, taking care of unsorted numbers in this bucket. In this case, the slaves are doing counting sort as the unsorted data arrive sequentially. A slave returns the sorted bucket to the master, who simply concatenates the results from all slaves together as a sorted array.

In our experiment, the unsorted array comprises one million integers, which are randomly distributed between 0 and 65535. We have all 12 hosts in our testbed working at the same time, with 1 master and 11 slaves for an individual sorting job.

All network flows are originally generated through the socket API provided by the official \texttt{Net} module. In order to test RepFlow and RepSYN provided by our {RepNet} module, all we need to do is to change the module \texttt{require} statements at the very beginning of the \texttt{node} script.

{\bf Mice Flows.}
The unsorted data distribution process from the master involves a large number of mice flows sending out to multiple destination slaves, because the unsorted numbers are scanned sequentially by the master. A buffering mechanism is used to reduce the flow fragmentation --- a chunk of unsorted numbers will not be sent out until a set of 20 numbers to the same destination slave is buffered. With buffering, these flows are still small in size ($<$1~KB).

{\bf Elephant Flows.}
When a slave completes its share of work, it returns the sorted results in the form of a large file to the master. We take these flows as elephants which will not be replicated by RepNet. 

{\bf Performance Metrics.}
In our experiment, each server is working as both master and slave at the same time. As a master node, it continuously generates new random unsorted data sets after the completion of the last job it coordinates. At the same time, it is working as a slave node for each one of the other 11 servers. In this case, the network traffic is a mixture of mice and elephant flows, whose communication pattern is much alike typical ones in a production cluster. Note that the starting time of each sorting master is delayed for several milliseconds randomly, in order to reduce flow concurrency at the beginning of our experiment.

We examine the CDF of the job completion times with different transport mechanisms, i.e. stack TCP, RepFlow and RepSYN. The timing of the job starts when the sorting master begins, i.e. starts reading the unsorted array from the input file, and stops as soon as the sorted array are successfully written to a text file.

\subsubsection{Job Completion Time Analysis}

\begin{figure}[t!]
	\vspace{-0cm}    
	\centering
	\includegraphics[width=0.4\textwidth]{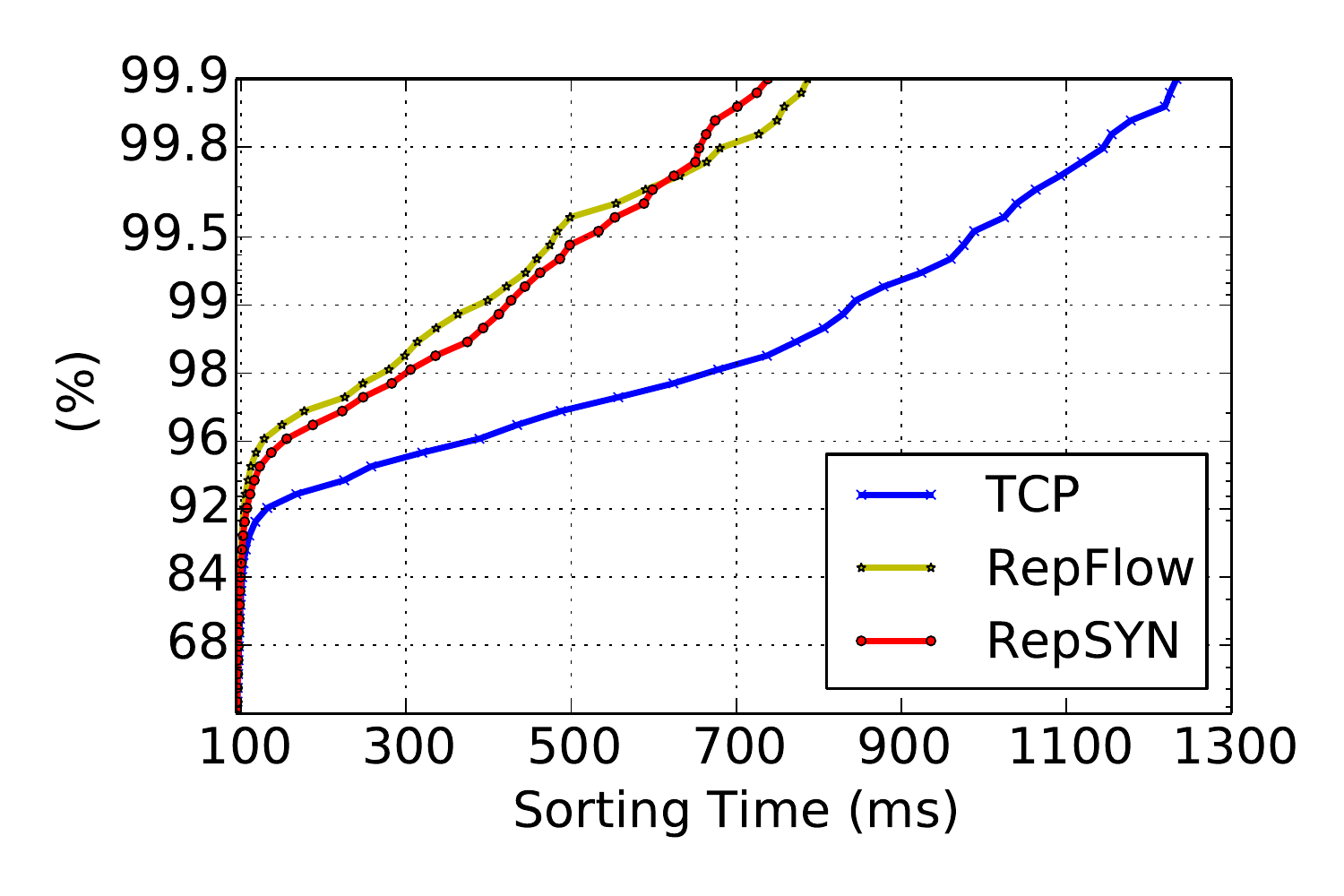}
	\vspace{-0.0cm} \caption{Job completion time CDF of the bucket sort application.}
	\label{fig:app-jct}
	\vspace{-0.4cm}
\end{figure}

We run the bucket sort application over 1,000 times on each machine with each transport mechanism, respectively. As a result, over 12,000 job completion times of similar sorting tasks are collected. The CDFs are plotted in Fig.~\ref{fig:app-jct}. Note that the y-axis is in log scale to emphasize the tail distribution.

Since bucket sort works in a partition-aggregation pattern, the job completion time will be determined by the last slave node to complete its assigned work. The long FCT of even one single flow may greatly degrade the application-level performance; therefore, the impact of the ``long tail'' of FCTs is magnified. Both of our implemented mechanisms, RepFlow and RepSYN, have shown outstanding ability to cut the long tail of the FCT distribution of mice flows, and Fig.~\ref{fig:app-jct} further highlights this benefit.

Since the network paths are idle in most of the time, most jobs ($\sim$85\%) can finish between 95 to 100~ms. 
However, due to the random occurrence of flash congestions, some of the jobs experience extremely long delay. With stack TCP, the 99.9th percentile job completion time can be as long as 1.2s, which is over 11x more than a job without congestion. By using {RepNet} module instead, both RepFlow and RepSYN are able to reduce the 99.9th percentile job completion time by $\sim$45\%, to 700--800~ms. The 99th percentile job completion time is reduced by $\sim$50\%.

To compare RepFlow and RepSYN, their CDF lines are similar with slight differences. RepFlow turns out to be a bit better in most (99.7\%) jobs, but has a much longer tail (nearly 100~ms) at the 99.9th percentile. The reason of this discrepancy is that the gathering sorting results may result in an incast pattern, with multiple-to-one elephant flow transmission. In most cases, these flows are not concurrent --- slaves typically do not finish their work at the same time, and RepFlow works smoothly. However when the elephant flows happen to have a high concurrency and incast happens, RepSYN is able to better survive the extreme cases.

\section{Mininet Emulation}
\begin{figure*}[ht!]
	\centering
	\begin{subfigure}[t]{0.32\textwidth}
		\centering
		\includegraphics[width=\textwidth]{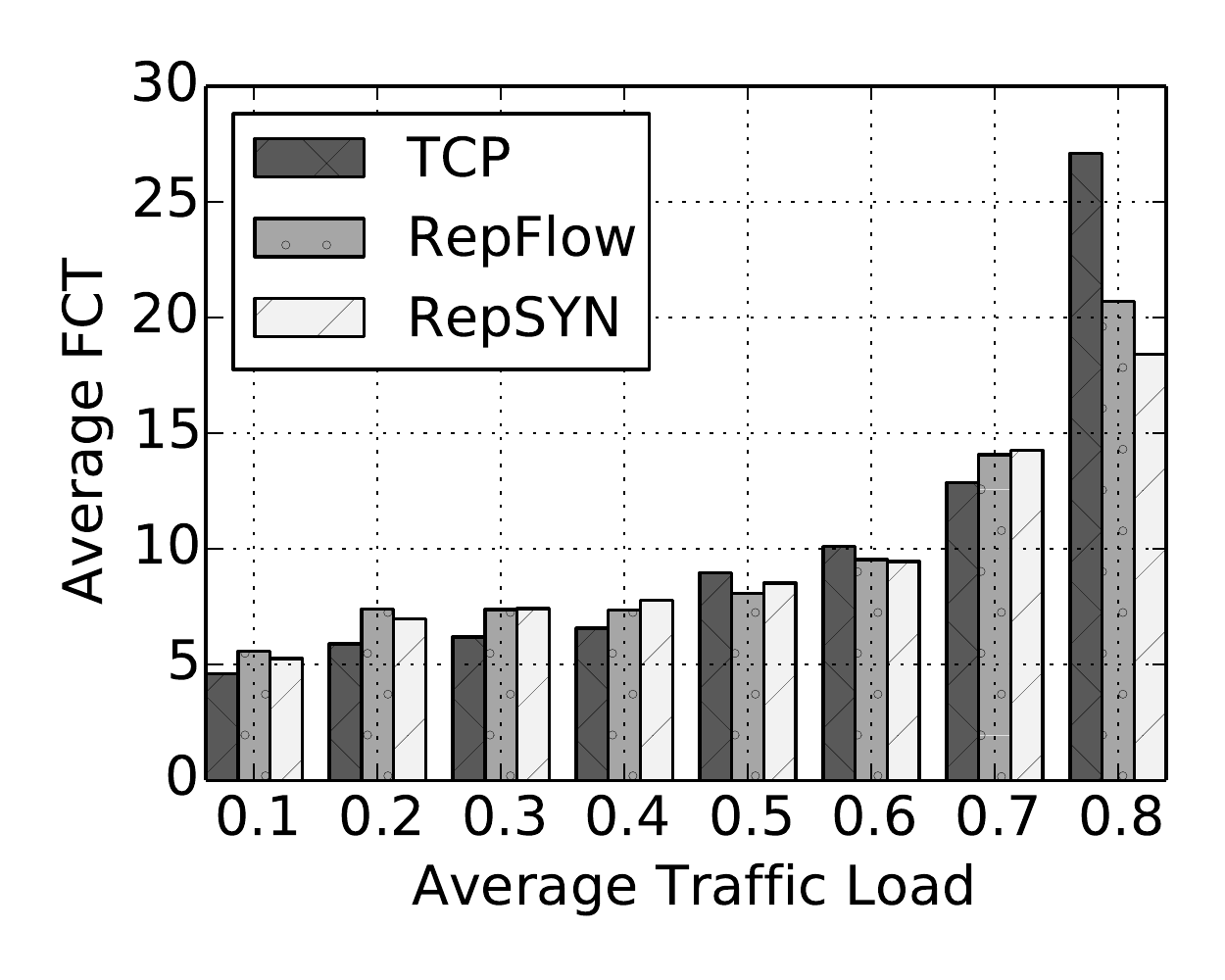}
	\end{subfigure}%
	~ 
	\begin{subfigure}[t]{0.32\textwidth}
		\centering
		\includegraphics[width=\textwidth]{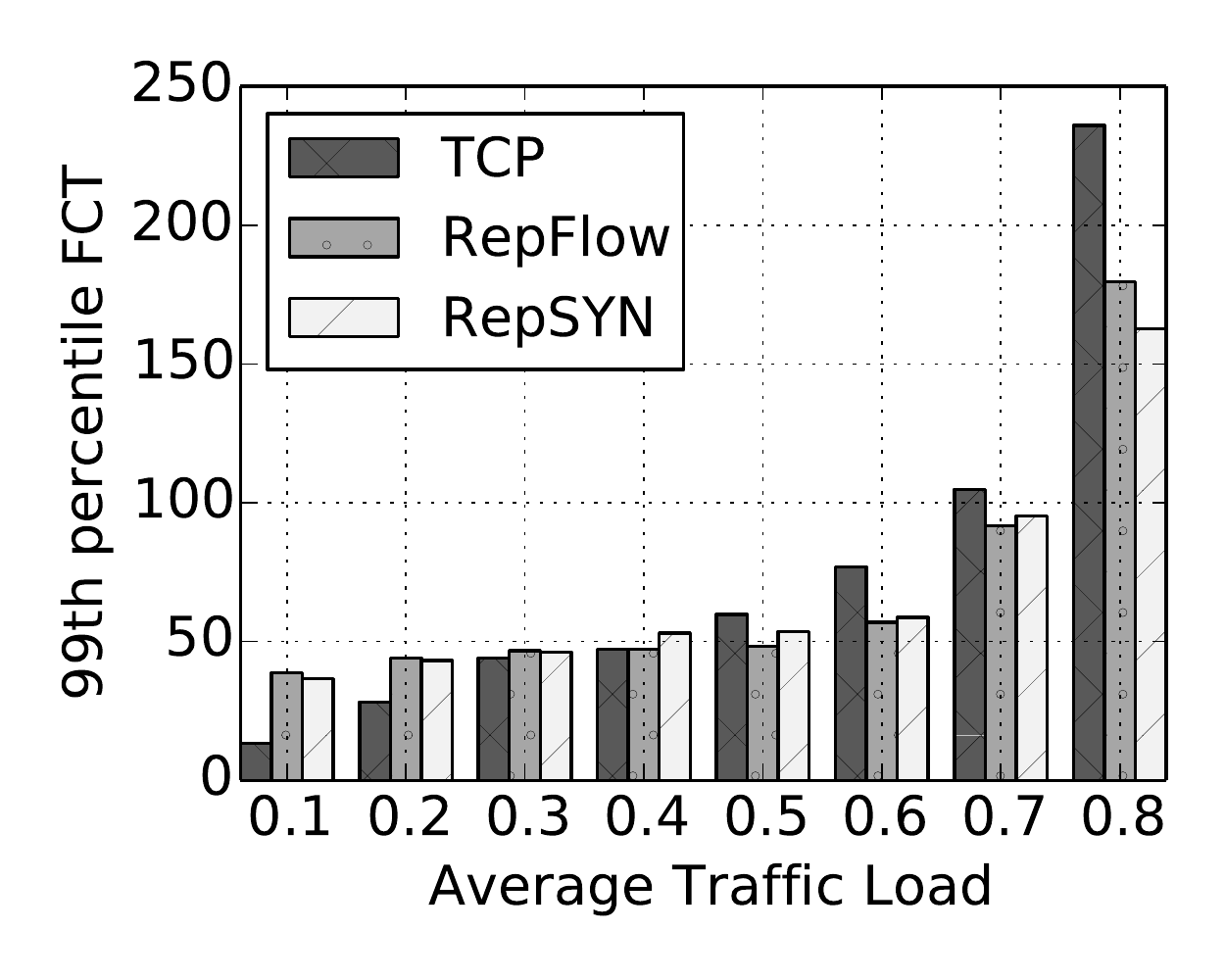}
	\end{subfigure}%
	~ 
	\begin{subfigure}[t]{0.32\textwidth}
		\centering
		\includegraphics[width=\textwidth]{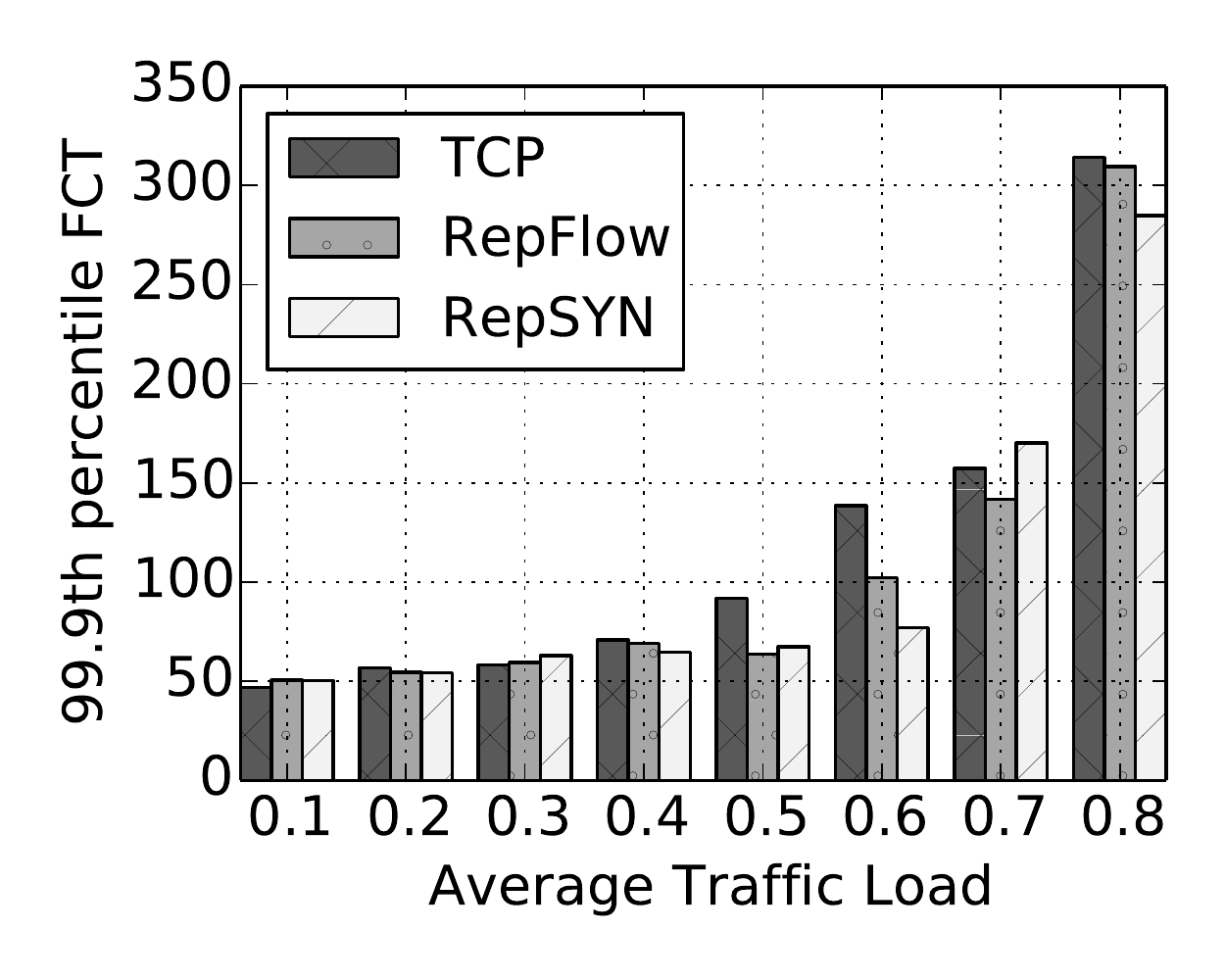}
	\end{subfigure}
	\vspace{-0.0cm} \caption{FCT comparison in Mininet with a fat-tree.}%
	\label{fig:mininet} \vspace{-0.2cm}
\end{figure*}

To verify the performance of {RepNet} in a relatively larger scale with higher path diversity, we conduct experiments using Mininet \cite{HHJL12}, with a 6-pod fat-tree and ECMP. Mininet is a high fidelity network emulator for software-defined networks on a single Linux machine. 
Routing is completely under the control of a centralized controller which, in our tests, is running on the same physical machine. All the scripts used for evaluation here is available online \cite{repnet_exp}.




\subsection{Mininet Configuration}

To guarantee high fidelity of the emulation results, we use a Mininet 2.2.0 virtual machine (official VM distributed with Ubuntu 14.04 LTS 64-bit) running on an Amazon EC2 \texttt{c3.4xlarge} instance, which has 16 vCPUs and 30GB memory available.

We create a 6-pod fat-tree without oversubscription. This is a 3-tier topology with 6 core switches, 18 aggregation switches, and 18 ToR switches. Each rack holds 3 hosts. As a result, it supports 54 hosts with up to 6 equal cost paths between two hosts for different pods. Note that all links in the topology are set to 50Mbps because of the limited switching ability on a single machine. The buffer size at each switch output port is configured to 100 packets.
To enable ECMP, an open-source POX controller module\footnote{\url{https://bitbucket.org/msharif/hedera/src}} is used. The controller implements the ECMP five-tuple hash routing as in RFC 2992.

\subsection{Emulation Results}

We plot the average, 99th percentile and 99.9th percentile FCT under various traffic loads in Fig.~\ref{fig:mininet}. The comparison methodology is similar to that in Fig.~\ref{fig:1NFCT} and Fig.~\ref{fig:2NFCT}, except that we use FCT rather than NFCT in this case.

{\bf Salient Benefit at Tail or High Load.}
Not surprisingly, both RepFlow and RepSYN show benefits for tail latency or under high load ($\geq 0.4$), and the figures show similar trends to Fig.~\ref{fig:1NFCT} and Fig.~\ref{fig:2NFCT}. However, one significant difference is that RepSYN is able to approximate or even outperform RepFlow. The reason is that with more paths available, congestion level on a single path is less fluctuating. Therefore, RTTs of the SYN packets can better estimate the congestion level throughout the transmission process of a single mice flow.

{\bf Low Traffic Load} ($\leq 0.4$).
However, under low loads, we cannot see much benefit from using RepNet. In some cases, they are even worse than the stack TCP. This is due to the controller overhead in Mininet which we explain now.

\subsection{Discussion} %
\label{sec:mininet-discussion}

Since Mininet is originally designed to emulate a software-defined network, all network traffic are controlled by a single centralized controller, i.e. a POX controller process in our experiment. This makes Mininet an imperfect tool for traditional network emulation.

When a flow initiates in Mininet, its SYN is identified as an unknown packet by the first switch it passes, and it is forwarded to the controller immediately. Then, the controller runs the ECMP routing algorithm for this packet, and installs new forwarding rules on all switches along the corresponding path. This process usually takes $\sim$1~ms (as the {ping} result suggests) even when the network is idle. With a large number of flows initiated around the same time the controller is easily congested.
Flow replication aggravates the controller overload. This results in the distortion of flow latency, which does not exist in real data center networks.

Nevertheless, in most cases, we can still benefit from using {RepNet} despite the controller overhead.

\section{Related Work}
\label{sec:related}


Low latency data center networking has been an active research area over the recent years. DCTCP \cite{AGMP10} and HULL \cite{AKEP12} use ECN-based congestion control to keep the switch queue occupancy low. DeTail \cite{ZDMB12}, DRB \cite{CXYG13}, and Expeditus \cite{WX14} design advanced multipath load balancing techniques to avoid congested paths for mice flows. D$^3$ \cite{WBKR11}, D2TCP \cite{VHV12}, and PDQ \cite{HCG12} use explicit deadline information for rate allocation, congestion control, and preemptive scheduling. DeTail \cite{ZDMB12} and pFabric \cite{AYSK13} present clean-slate designs of the entire network fabric that prioritize latency sensitive short flows, while PIAS \cite{BCCH14} presents a flow size agnostic priority scheduling mechanism. All of them require modifications to switches and end-hosts. Note priority queueing is widely supported in commodity switches and in principle can be used to expedite mice flows. However this will interact negatively with its existing use in traffic differentiation based on applications and purposes\footnote{Packets from control protocols typically have higher priority. Also, traffic of production systems have higher priority than traffic for experimental and development purposes.}, which is fairly common in production networks.

We also comment that the general idea of using replication to improve latency has gained increasing attention. Mitzenmacher's seminal work on ``power of two choices'' \cite{M96} proposes for a request to randomly sample two servers, and queue at the one with less queueing to achieve good load balancing. Google reportedly uses request replication to rein in the tail response times \cite{D12}. Vulimiri et al. \cite{VGMS13} argue for the use of redundant operations to improve latency in various systems. To our best knowledge, our work is among the first to provide a readily deployable implementation and testbed evaluation of flow replication in data centers.

\section{Concluding Remarks}

We presented the design, implementation, and evaluation of \texttt{RepNet}, a low-latency application layer transport module based on {\tt node} which provides socket APIs to enable flow replication. Experimental evaluation on a real testbed and in Mininet demonstrates its effectiveness on both mean and tail latency for mice flows. We also proposed RepSYN to alleviate its performance degradation in incast scenarios. 

An interesting observation inspired by this work is that efficient multipath routing in data center networks is worth further investigation. Because randomly choosing another path indeed makes a difference in RepNet, a congestion-aware routing algorithm that aims to choose the best path may provide even more latency reduction, though this imposes daunting challenges of collecting global and timely congestion information in a microsecond time scale \cite{AEDV14,WX14}. 

\section*{Acknowledgments}
We thank Fred Baker from Cisco, Shuang Yang from Google, and Baochun Li from University of Toronto for helpful feedback and suggestions. 
\bibliographystyle{abbrv}
\bibliography{IEEEabrv,main}

\end{document}